\documentclass[a4paper,11pt]{article}
\pdfoutput=1 

\usepackage{jcappub} 

\usepackage[T1]{fontenc} 
\usepackage{xspace}
\usepackage{multirow}
\usepackage{booktabs}
\usepackage{graphicx}
\usepackage{subcaption}
\usepackage{booktabs}   
\usepackage{array}      
\usepackage{makecell}   
\usepackage{cleveref}

\newcolumntype{C}[1]{>{\centering\arraybackslash}m{#1}}

\usepackage{cellspace}
\usepackage{xcolor}
\definecolor{darkblue}{rgb}{0.0, 0.0, 0.55}
\definecolor{darkgreen}{rgb}{0.0, 0.55, 0.2}
\definecolor{darkred}{rgb}{0.55, 0.0, 0}

\title{Weighing neutrinos with 21cm Intensity Mapping at the SKAO}

\author[a,b,c]{G. Autieri,}
\author[d]{M. Berti,}
\author[e,f]{M. Spinelli,}
\author[a,b,c]{B. S. Haridasu,}
\author[a,b,c,g,h]{M. Viel}

\affiliation[a]{SISSA- International School for Advanced Studies, Via Bonomea 265, 34136 Trieste, Italy}
\affiliation[b]{INFN – National Institute for Nuclear Physics, Via Valerio 2, I-34127 Trieste, Italy}
\affiliation[c]{IFPU, Institute for Fundamental Physics of the Universe, via Beirut2, 34151 Trieste, Italy}
\affiliation[d]{Département de Physique Théorique and Center for Astroparticle Physics, Université de Genève, Quai E. Ansermet 24, CH-1211 Genève 4, Switzerland}
\affiliation[e]{Observatoire de la C\^{o}te d’Azur, Laboratoire Lagrange, Bd de l’Observatoire, CS 34229, 06304 Nice cedex 4, France}
\affiliation[f]{Department of Physics and Astronomy, University of the Western Cape, Robert Sobukwe Road, Cape Town 7535, South Africa}
\affiliation[g]{INAF, Osservatorio Astronomico di Trieste, Via G. B. Tiepolo 11, I-34131 Trieste, Italy}
\affiliation[h]{Italian Research Center on High Performance Computing, Big Data and Quantum Computing}

\emailAdd{gautieri@sissa.it}
\emailAdd{maria.berti@unige.ch}

\abstract{We explore the constraining power of future 21cm intensity mapping (IM) observations at the SKAO, focusing primarily on the sum of neutrino masses, $\Sigma m_\nu$. We forecast observations of the 21cm IM auto-power spectrum as well as the 21cm IM and galaxy surveys cross-correlation power spectrum. We construct different synthetic data sets of observations for the 21cm IM observables in the redshift range $0 < z < 3$. For galaxy clustering, we consider two stage-IV surveys to mimic a DESI-like and Euclid-like cross-correlation signal. 
We study the impact of assuming three different fiducial values for the sum of neutrino masses, i.e. $\Sigma m_\nu = 0.06, 0.1, 0.4$ eV, in the synthetic data sets. To investigate the constraining power of the forecasted 21cm observations, we build a likelihood code that will be made publicly available upon publication. The results of the analysis, obtained through Markov Chain Monte Carlo techniques, are promising. We find that the 21cm auto-power spectrum alone could provide an upper limit on the sum of neutrino masses of $\Sigma m_\nu < 0.287$ eV, at $95\%$ confidence level, for the case of the lowest fiducial value of $\Sigma m_\nu$. This result is comparable to the upper limits provided by cosmic microwave background (CMB) observations alone. When combining the 21cm auto-power spectrum synthetic data set with Planck 2018 CMB measurements, we find a tighter upper limit of $\Sigma m_\nu < 0.105$ eV, which improves on the constraints from Planck alone. We obtain a similar result already at the level of 21cm and galaxy clustering cross-correlation power spectrum, whose detection is more easily achieved as they are less affected by systematic effects. Combining synthetic data sets with Planck 2018 data, we find the upper limits of $\Sigma m_\nu < 0.116$ eV and $\Sigma m_\nu < 0.117$ eV for the 21cm signal in cross-correlation with the DESI-like and Euclid-like surveys, respectively. These constraints are comparable to those obtained by combining Planck data with the 21cm auto-power spectrum synthetic data sets, thus supporting the case for 21cm cross-correlation detections. 
}

\begin{document}
\maketitle
\flushbottom

\section{Introduction}
\label{sec:intro}
Neutrinos are one of the most puzzling particles in the Standard Model of particle physics. They were first postulated by Pauli in 1930 in order to address the apparent violation of energy conservation in $\beta$-decays. They were later detected for the first time by Cowan and Reines in 1956 \citep{Cowan:1956rrn}. Neutrinos are known to come in three different flavours, $\nu_e, \, \nu_\mu, \, \nu_\tau$ and were initially considered to be massless particles. However, observations of the so-called neutrino oscillations, i.e. the change in flavour of neutrinos, have confirmed that neutrinos are massive. Measurements of neutrino oscillations from lab experiments allow to constrain the squared mass difference between different neutrino eigenstates: $\Delta m^2_{12} = 7.5 \times 10^{-5}\,\mathrm{eV}^2$ and $\vert\Delta m^2_{31}\vert = 2.55\times 10^{-3}\,\mathrm{eV}^2$ \citep{Esteban:2020cvm, deSalas:2020pgw}. Instead, $\beta$-decay experiments offer a direct measurement of the neutrino mass. The latest results from the KATRIN collaboration, obtained by measuring the tritium $\beta$-decay spectrum, placed an upper limit on the effective mass of $m_{\mathrm{\beta}} < 0.45$ eV $(90\%$ CL) \citep{KATRIN:2021uub, Katrin:2024tvg}.

Other than particle physics experiments, cosmology offers the biggest lab in which to test neutrinos and measure their mass. In the $\Lambda$CDM model, massive neutrinos are relativistic in the early Universe and become non-relativistic at later times, influencing both the background evolution of the Universe and the growth of structures on intermediate and small scales. Since the expansion history of the Universe is reflected in the positions of the Baryon Acoustic Oscillation (BAO) peaks, BAO can be used to constrain $\Sigma m_\mathrm{\nu}$ \citep{2012PhRvD..86j3518P, Vagnozzi:2017ovm, Loureiro:2018pdz, RoyChoudhury:2019hls, Palanque-Delabrouille:2019iyz, DiValentino:2021hoh, Brieden:2022lsd,DESI:2024mwx, DESI:2024hhd, DESI:2025zgx}. Specifically, relativistic neutrinos have large thermal velocities, which allow them to free-stream out of high-density regions, damping perturbations on scales smaller than the free-streaming length. This leads to a suppression of the matter power spectra on small scales \citep{Lesgourgues:2006nd, Wong:2011ip, TopicalConvenersKNAbazajianJECarlstromATLee:2013bxd}. This imprint has been widely exploited as a way to constrain the sum of neutrino masses through measurements of biased tracers of matter distribution alone and in combination with Cosmic Microwave Background (CMB) data  \citep{planck:2018,Elgaroy:2002bi, SDSS:2003eyi, WMAP:2003elm, WMAP:2003ivt, SDSS:2004kqt, Goobar:2006xz, Seljak:2006bg, Viel:2010bn, Palanque-Delabrouille:2015pga, Rossi:2014nea, Yeche:2017upn}. In the near future, next-generation stage-IV surveys, such as the Dark Energy Spectroscopic Instrument (DESI) \citep{DESI:2016fyo, DESI:2024uvr} and Euclid \citep{Euclid:2019clj, Euclid:2024imf}, will improve the current constraints on neutrino masses by probing larger cosmic volumes and extending the observed redshift range beyond the current limits. However, to achieve the highest precision possible and deal with possible hidden systematics, it will be crucial to combine these data with complementary observations of the large-scale structures. To this end, a promising and innovative probe is the intensity mapping (IM) of the 21cm neutral hydrogen line \citep{Bharadwaj:2000, Battye:2004, Loeb:2008hg, Kovetz:2017}. The main advantage of IM observations is the possibility of efficiently probing large volumes at the expense of a relatively poor angular resolution. The large sky area of these observations offers significant potential for detecting the HI signal in cross-correlation with galaxy surveys. 

One of the most promising upcoming experiments in this area is the SKAO\footnote{See \url{https://www.skao.int}.}. The SKAO, which is currently under construction, will consist of the SKA-Low and SKA-Mid telescopes, located in Australia and South Africa, respectively. IM surveys that will aim to detect the 21cm signal at cosmological scales up to redshift $z\sim 3$ using the SKA-Mid telescope have been proposed. \citep{Bacon:2018, Battye:2012tg, Santos:2015}. At these redshifts, the 21cm signal acts as a biased tracer of the underlying matter field, making it possible to detect the imprint of massive neutrinos on the large-scale clustering properties of matter. Additionally, IM measurements from the SKAO will enable the detection of the 21cm signal in cross-correlation with current and future galaxy surveys carried out by state-of-the-art telescopes, such as DESI and Euclid. This is of the utmost importance, as detections of the 21cm IM signal in cross-correlations are expected to be more robust against instrumental noise and systematics than the 21cm auto-correlation. This will enable cross-correlation data to be available sooner with respect to auto-correlation power spectrum detections. Notably, the SKA-Mid precursor, MeerKAT, has been conducting IM surveys for cosmology, reporting a first detection of the HI signal in cross-correlation with WiggleZ \citep{Cunnington:2022uzo, 2010MNRAS.401.1429D, 2018MNRAS.474.4151D} and GAMA \citep{MeerKLASS:2024ypg, 2009A&G....50e..12D, 2011MNRAS.413..971D} galaxies.

In this work, we explore the constraining power on the neutrino mass of 21cm IM with the SKAO. To achieve this, we build upon the formalism developed in \citep{Berti:2021ccw, Berti:2022ilk, Berti:2023viz}, in which the authors explored the constraining power of forecasted SKAO 21cm IM observations on the cosmological parameters. We extend the formalism and the analysis pipeline to cosmologies with massive neutrinos, and we build synthetic data sets of 21cm IM and 21cm-galaxy survey cross-correlation observations, simulating realistic future observations with SKA-Mid. In the construction of synthetic data sets, we consider several fiducial values for the sum of neutrino masses, $\Sigma m_\nu = 0.06,\; 0.1,\; 0.4$ eV. These values are motivated by current constraints on the neutrino mass. The value of $\Sigma m_\nu = 0.06$ eV agrees with the lower bounds imposed by particle physics experiments and the most recent upper limits of $\Sigma m_\nu < 0.0642$ eV found with CMB and DESI BAO most recent observations~\cite{DESI:2025zgx}. The choice of the other two values mimics more extreme scenarios. We investigate the implications for 21cm observations to favor a slightly higher value of the neutrino mass (0.1 eV), not in agreement with current DESI constraints, but still compatible with CMB alone bounds ($\Sigma m_\nu < 0.26$ eV)~\cite{planck:2018}. Instead, the value of 0.4 eV, not compatible with current constraints, allows us to explore an extreme scenario and validate our pipeline.  We then conduct a Bayesian analysis to derive constraints $\Sigma m_\mathrm{\nu}$ and the cosmological parameters describing the $\Lambda$CDM model. Additionally, we combine our data sets with Planck 2018 CMB data to investigate the synergies between early and late Universe probes. The likelihood code used in our analysis will be made publicly available upon publication.

We note that forecasts on the constraining power of 21cm IM observations on the neutrino mass with the SKAO have been presented in previous works, e.g. \cite{Villaescusa-Navarro:2015cca}. Here we highlight the key differences with respect to our work: \textit{i}) in \cite{Villaescusa-Navarro:2015cca}, they combine a SKA-Low survey with a SKA-Mid survey, probing the redshift range $0 < z < 6$, which is wider than the range $0 < z < 3$ that we consider in our work, \textit{ii}) in \cite{Villaescusa-Navarro:2015cca} the authors use a hydrodynamical simulations-based approach while, in this work, we make use of analytical models, \textit{iii}) we conduct an MCMC analysis to estimate the full posterior distributions, as opposed to the Fisher matrix formalism employed in \cite{Villaescusa-Navarro:2015cca}, \textit{iv}) we extend the analysis to forecasted 21cm IM signal detections in cross-correlation with galaxy surveys such as DESI and Euclid.

This paper is organized as follows: in \autoref{sec:methods} we discuss the adopted methodology. Specifically, in \autoref{sec:modelling_21cm} we discuss the impact of massive neutrinos on the theoretical power spectra modeling for both 21cm IM and 21cm-galaxy clustering cross-correlation. In \autoref{sec:tomographic_data_set} we detail the construction of the synthetic data sets. The numerical analysis is discussed in \autoref{sec:numerical_analysis}. In \autoref{sec:results} we discuss the results, presenting forecasted constraints from 21cm IM alone and in combination with Planck 2018 data in \autoref{sec:21cm results} and forecasted constraints from 21cm-galaxy clustering cross-correlation data alone and in combination with Planck 2018 in \autoref{sec:cross_results}. A summary of the results is presented in \autoref{sec:conclusion}.

\section{Methods}
\label{sec:methods}
In this work, we build upon the formalism and framework developed in \citep{Berti:2023viz,Berti:2022ilk, Berti:2021ccw}. In \autoref{sec:modelling_21cm}, we extend the formalism to cosmologies with different neutrino masses. In \autoref{sec:tomographic_data_set}, we describe the construction of the synthetic tomographic data set of SKA-Mid observations and the construction of the synthetic data sets of cross-correlation measurements with future surveys. Details on the numerical analysis are discussed in \autoref{sec:numerical_analysis}.

\subsection{Modeling power spectra for massive neutrino cosmologies}
\label{sec:modelling_21cm}
Massive neutrinos influence structure formation in two ways \citep{Lesgourgues:2006nd, Wong:2011ip, TopicalConvenersKNAbazajianJECarlstromATLee:2013bxd}. Firstly, they impact the background evolution of the universe. Early on, massive neutrinos behave like radiation but later transition to the non-relativistic regime, altering the evolution of the Universe through the Friedmann equations. The second effect occurs after neutrinos decouple. At this stage, they free-stream out of high-density regions, suppressing perturbations on scales smaller than the free-streaming scale. This suppression reduces the amplitude of the matter power spectrum on small scales, providing a way to constrain neutrino masses. The free-streaming scale is inversely related to neutrino mass, so higher masses result in smaller free-streaming scales, shifting the effect to smaller scales. Previous studies \citep{Villaescusa-Navarro:2013pva,Castorina:2013wga} have shown that neutrinos do not contribute significantly to the CDM halo mass. Consequently, to investigate the influence of massive neutrinos on the large-scale structures, we can neglect their contribution to clustering quantities and define the model of the non-linear 21cm power spectrum as~\citep{Villaescusa-Navarro:2015cca, Bacon:2018}
\begin{equation}
    P_{21}(z,k,\mu) = \overline{T}_\mathrm{b}^2(z) \Big[b_{\mathrm{HI}}(z) + f_{\mathrm{CDM + b}}(z, k) \,\mu^2 \Big]^2\, P_{\mathrm{CDM + b}}(z,k) + P_{\mathrm{SN}}(z),
    \label{21cm_ps}
\end{equation}
where $\overline{T}_\mathrm{b}(z)$ is the HI brightness temperature, $b_{\mathrm{HI}}$ is the HI bias, $f_{\mathrm{CDM + b}}$ is the growth rate of CDM and baryons, $\mu = \hat{k}\cdot \hat{z}$ is the cosine of the angle between the wavevector $k$ and the line-of-sight, $ P_{\mathrm{CDM + b}}$ is the non-linear power spectrum, and $P_{\mathrm{SN}}(z)$ is the shot noise term. Both the non-linear matter power spectrum $P_{\mathrm{CDM + b}}(z,k)$ and the growth rate $f_{\mathrm{CDM + b}}(z, k)$ are computed considering only the contributions from CDM and baryons. The matter power spectrum is computed with the Einstein-Boltzmann solver \texttt{CAMB} \citep{Lewis:2000}\footnote{See \href{https://camb.info/}{CAMB webpage} for more. We note that the non-linear corrections to the matter power spectrum are computed by means of \texttt{HMcode-2020} \citep{Mead:2020vgs}.}, while the growth rate is computed numerically as
\begin{equation}
    f(z, k) = -( 1 + z) \,\frac{d\log{D(z, k)}}{dz},
    \label{f_formula}
\end{equation}
where $D(k, z)$ is the growth factor, derived from the matter power spectrum as $P(z, k) = D^2(z, k) P(0, k)$. It is important to note that the growth factor $D(z, k)$, and consequently also $f(z, k)$, depend on the wavenumber $k$ due to the free-streaming of massive neutrinos (see e.g.~\citep{Castorina:2013wga}). For the redshift evolution of the brightness temperature, we employ the parametrization of \citep{Battye:2012tg}. For $b_{\mathrm{HI}}(z)$ and $P_{\mathrm{SN}}(z)$ we use interpolated numerical results from hydrodynamical simulations \citep{Villaescusa-Navarro:2014rra,Villaescusa-Navarro:2018}.
To more accurately mimic a real-world observation we introduce an exponential factor that approximates the smoothing of perpendicular modes due to the effect of a Gaussian telescope beam \citep{Battye:2012tg,Cunnington:2020mnn,Cunnington:2022ryj,Soares:2020zaq} which can be written in terms of the physical dimension of the beam, $R_{\mathrm{beam}}$, as
\begin{equation}
    B(z,k,\mu) = \exp{\bigg[-\frac{k^2 R^2_{\mathrm{beam}}(z)}{2}\,(1 - \mu^2)\bigg]}.
    \label{beam_suppression}
\end{equation}
Moreover, we incorporate in our model the Alcock-Paczynski (AP) effects
\citep{Alcock:1979mp}, which arise from anisotropies
caused by discrepancies between the assumed fiducial cosmology and the true underlying cosmology. In order to model these anisotropies in the transverse and radial directions, we introduce the AP parameters
\begin{equation}
    \alpha_{\perp}(z) = \frac{D_{\mathrm{A}}}{D_{\mathrm{A}}^{\mathrm{fid}}}
    \qquad \mathrm{and} \qquad \alpha_{\parallel}(z) = \frac{H^{\mathrm{fid}}(z)}{H(z)},
    \label{AP_params}
\end{equation}
where $D_{\mathrm{A}}^{\mathrm{fid}}(z)$ and $H^{\mathrm{fid}}(z)$ are the fiducial values of the angular diameter distance and the Hubble parameter at redshift $z$, respectively. Then, the true wavenumber and cosine of the angle between the wavevector and the line-of-sight, $q$ and $\nu$ respectively, can be related to the observed ones, $k$ and $\mu$, as
\begin{equation}
    q = \frac{k}{\alpha_{\perp}}\sqrt{1 + \mu^2 \biggl(\frac{\alpha^2_{\perp}}{\alpha^2_{\parallel}} - 1\biggr)}
\end{equation}
and
\begin{equation}
    \nu = \frac{\alpha_{\perp}\mu}{\alpha_{\parallel}\sqrt{1 + \mu^2\biggl(\frac{\alpha^2_{\perp}}{\alpha^2_{\parallel}} - 1\biggr)}}.
\end{equation}
Taking into account these effects, we can then model the observed 21cm power spectrum as
\begin{equation}
    \hat{P}_{21}(z,k,\mu) = \frac{1}{\alpha_{\parallel}\alpha^2_{\perp}} B^2(z,q,\nu) P_{21}(z, q, \nu).
    \label{ps21_observed}
\end{equation}
Similarly to what we did above for the 21cm power spectrum, the model for the cross-correlation power spectrum between the 21cm signal and galaxy clustering can be extended to different massive neutrino cosmologies as
\begin{equation}
    P_{\mathrm{21,g}}(z,k,\mu)=\overline{T}_{\mathrm{b}}(z)\Bigl(b_{\mathrm{HI}}(z) + f_{\mathrm{CDM + b}}(k, z)\mu^2\Bigr)\Bigl(b_{\mathrm{g}}(z) + f_{\mathrm{CDM + b}}(k, z)\mu^2\Bigr)\, P_{\mathrm{CDM + b}}(z,k,\mu),
    \label{cross_PS}
\end{equation}
where $b_\mathrm{g}$ is the galaxy bias and all other quantities have been defined above. Including the AP effect and the beam smoothing, we can model the observed cross-correlation signal as \footnote{When working with real data, a coefficient $r$ is introduced to account for additional unknown effects that may change the estimate of the correlation degree \citep{Cunnington:2022ryj}. In this work, we use a fiducial value of $r = 1$ to construct the synthetic data sets. Possible variations of $r$ are intrinsically already taken into account when varying the nuisance parameters. For this reason, we do not explicitly include $r$ in the model.}

\begin{equation}
    \hat{P}_{\mathrm{21,g}}(z,k,\mu) = \frac{1}{\alpha_{\parallel}\alpha^2_{\perp}}\,B(z,q,\nu)\,P_{\mathrm{21,g}}(z,q,\nu).
    \label{PS_cross_obs}
\end{equation}
The observed power spectra are anisotropic and they can be decomposed using Legendre polynomials $\mathcal{L}_\ell (\mu)$. The coefficients of the expansion, i.e. the power spectrum multipoles, are given by
\begin{equation}
    P_{X,\ell}(z,k) = \frac{2\ell +1}{2} \int_{-1}^{1} \mathrm{d}\mu\,\mathcal{L}_\ell (\mu)\,\hat{P}_X(z,k,\mu),
    \label{multipoles}
\end{equation}
where $X$ can be either $X = 21$ or $X = 21,g$ to indicate the 21cm auto or cross power spectrum, respectively. In this work, we will forecast observations of both the monopole ($\ell = 0$) and quadrupole ($\ell = 2$), for which, $\mathcal{L}_0(\mu) = 1$ and, $\mathcal{L}_2(\mu) = \frac{1}{2}(3\mu^2 -1)$, respectively.
\subsection{Forecasted 21cm observables synthetic data sets}
\label{sec:tomographic_data_set}
In this section, we present the recipe we follow to construct the synthetic data sets for the 21cm observables. The forecasted 21cm auto-power spectrum observations are presented in \autoref{sec:21cm_synt_auto}, while the ones for 21cm and galaxy sclustering cross-correlation power spectrum are discussed in \autoref{sec:modeling cross}.

\subsubsection{21cm intensity mapping auto-power spectrum}
\label{sec:21cm_synt_auto}
We focus on the HI IM signal that can be measured with the SKAO. Following \citep{Bacon:2018}, we consider a combination of two different surveys: a Medium-Deep Band 2 survey in the frequency range $0.95\, - 1.75\;$ GHz, that is in the redshift range $0\, -\, 0.5$, covering a sky area of $5000\;\mathrm{deg}^2$, and a Wide Band 1 survey in the frequency range $0.35\, - 1.05\;$ GHz (corresponding to the redshift range $0.35\, - 3$) covering a sky area of $20000\;\mathrm{deg}^2$. We construct synthetic observations for six equally-spaced, non-overlapping redshift bins with a width of $\Delta z = 0.5$ and centered at redshifts $z_c = \{0.25, 0.75, 1.25, 1.75, 2.25, 2.75\}$.

For each redshift bin, the survey specifications define the volume of the observations and therefore fix the range of accessible scales. The largest scale accessible for each redshift bin is, in Fourier space, defined by the observed volume
\begin{equation}
    k_{\mathrm{min}}(z_c) = \frac{2\pi}{\sqrt[3]{V_{\mathrm{bin}}(z_c)}},
    \label{kmin}
\end{equation}
where $V_{\mathrm{bin}}(z_c)$ is the volume of each redshift bin and it is computed as 
\begin{align}
    V_{\mathrm{bin}}(z_c) &= \Omega_{\mathrm{sur}}\int_{z_c - \Delta z/2}^{z_c + \Delta z/2}\, \mathrm{d}z^{\prime}\,\frac{\mathrm{d}V}{\mathrm{d}z^{\prime}\mathrm{d}\Omega} \\
    &= \Omega_{\mathrm{sur}}\int_{z_c - \Delta z/2}^{z_c + \Delta z/2}\, \mathrm{d}z^{\prime}\,\frac{c r(z^{\prime})^2}{H(z^{\prime})},
    \label{Vbin}
\end{align}
with $r(z)$ the comoving distance and $\Omega_{\mathrm{sur}}$ is the survey volume.\\
The smallest available scale is instead fixed by the damping due to the telescope beam and it can be estimated as 
\begin{equation}
    k_{\mathrm{max}}(z_c) = \frac{2\pi}{R_{\mathrm{beam}}(z_c)}.
    \label{kmax}
\end{equation}
We also choose a fixed $k-$bin width $\Delta k(z_c)$ large enough so that $k-$modes can be treated as independent, that is $\Delta k(z_c) = 2 k_{\mathrm{min}}(z_c)$. We can now compute the number of modes per $k$ and $\mu$ bins at a given redshift, $N_{\mathrm{modes}}(z_c, k, \mu)$ as
\begin{equation}
    N_{\mathrm{modes}}(z_c, k, \mu) =  \frac{k^2 \Delta k (z_c) \Delta \mu(z_c)}{8\pi^2}\,V_{\mathrm{bin}}(z_c),
    \label{Nmodes_1}
\end{equation}
where $\Delta \mu(z_c)$ is the $\mu$ bin width. However, in our analysis, we focus on the power spectra multipoles, thus, we integrate over all possible values of $\mu$ in the interval $\mu \in (-1, 1)$. In our case, then, the number of modes reduces to
\begin{equation}
    N_{\mathrm{modes}}(z,k) = \frac{k^2 \Delta k(z_c)}{4\pi^2}V_{\mathrm{bin}}(z_c).
    \label{Nmodes_2}
\end{equation}
In order to estimate the errors on the power spectrum measurements, we follow \citep{Bernal:2019jdo} and compute the variance per $k$ and $\mu$ bin as
\begin{equation}
    \sigma_{21}(z, k ,\mu) = \frac{\hat{P}_{21}(z,k,\mu) + P_{\mathrm{N}}(z)}{\sqrt{N_{\mathrm{modes}}(z,k,\mu)}},
    \label{sigma21}
\end{equation}
where $P_{\mathrm{N}}(z)$ is the noise power spectrum that, for a single-dish IM SKAO-like experiment, can be modeled as \citep{bull:2015, Bernal:2019jdo}
\begin{equation}
    P_{\mathrm{N}}(z) = \frac{T^2_{\mathrm{sys}}\, 4\pi f_{\mathrm{sky}}}{N_{\mathrm{dish}}\, t_{\mathrm{obs}}\,\delta\nu}\frac{V_{\mathrm{bin}}(z)}{\Omega_{\mathrm{sur}}},
    \label{Pnoise}
\end{equation}
where $T^2_{\mathrm{sys}}$ is the system temperature, $f_{\mathrm{sky}}$ is the fraction of the sky covered by the survey, $N_{\mathrm{dish}}$ is the number of dishes and $t_{\mathrm{obs}}$ is the observing time. As stated above, for the HI IM analysis we use the 21cm power spectrum monopole and quadrupole. Thus, to compute the errors on the multipoles, we define the covariance matrix between multipoles $\ell$ and $\ell^{\prime}$ as \citep{Bernal:2019jdo}
\begin{equation}
    C_{\ell,\ell^{\prime}}(z,k) = \frac{(2\ell + 1)(2\ell^{\prime}+1)}{2}\,\int_{-1}^{1}\,\mathrm{d}\mu\, \mathcal{L}_{\ell}(\mu)\,\mathcal{L}_{\ell^{\prime}}(\mu)\,\sigma^2(z,k,\mu).
    \label{multipoles_covariance}
\end{equation}
Therefore, the error on each point in a given data set can be computed as
\begin{equation}
    \sigma_{\hat{P_{\ell}}}(z,k) = \sqrt{C_{\ell \ell}(z,k)}.
    \label{sigma_multipoles}
\end{equation}
Equation \eqref{multipoles_covariance} can also be used to estimate the covariance between different multipoles, therefore, the most general covariance matrix for monopole and quadrupole at a fixed redshift is a block matrix of the form
\begin{equation}
C(z) = 
    \begin{pmatrix}
        C_{00}(z) &\; C_{02}(z)\\
        C_{20}(z) &\; C_{22}(z)
    \end{pmatrix}, 
    \label{block_form_cov}
\end{equation}
where each block is a diagonal matrix whose entries are computed with Equation \eqref{multipoles_covariance}.
\begin{table}
	\centering
	\begin{tabular}{lccc}
 \toprule
 Parameters & $\Sigma m_\nu^{\rm fid} =0.06$  & $\Sigma m_\nu^{\rm fid} =0.1$   &  $\Sigma m_\nu^{\rm fid} = 0.4$\\ 
  \midrule
   $\Omega_b h^2$ & $0.02238$ & $0.02235$ & $0.02215$\\
   $\Omega_c h^2$ & $0.11987$ & $0.12032$ & $0.12209$\\
   $n_\mathrm{s}$ & $0.96585$ & $0.96437$ & $0.95916$\\
   $\ln (10^{10} A_s)$ & $3.0444$ & $3.0426$ & $3.0529$\\
   $\tau$ & $0.0543$ & $0.05326$ & $0.0561$\\
   $100 \theta_{\rm MC}$ & $1.04091$ & $1.04082$ & $1.04049$\\
   $H_0\;$ [km/s/Mpc] & $67.41$ & $66.91$ & $63.43$\\
   $\Omega_{\rm m}$ & $0.3144$ & $0.3210$ & $0.3692$\\
   \bottomrule
	\end{tabular}
    
 	\caption{Fiducial values for the cosmological parameters from Planck 2018 \citep{planck:2018} best-fit results for different fixed values of the total neutrino mass.}
    \label{tab:fiducial_cosmos}
\end{table}
\begin{figure}
    \centering
    \begin{subfigure}{1.0\textwidth}
        \includegraphics[width=1.0\textwidth]{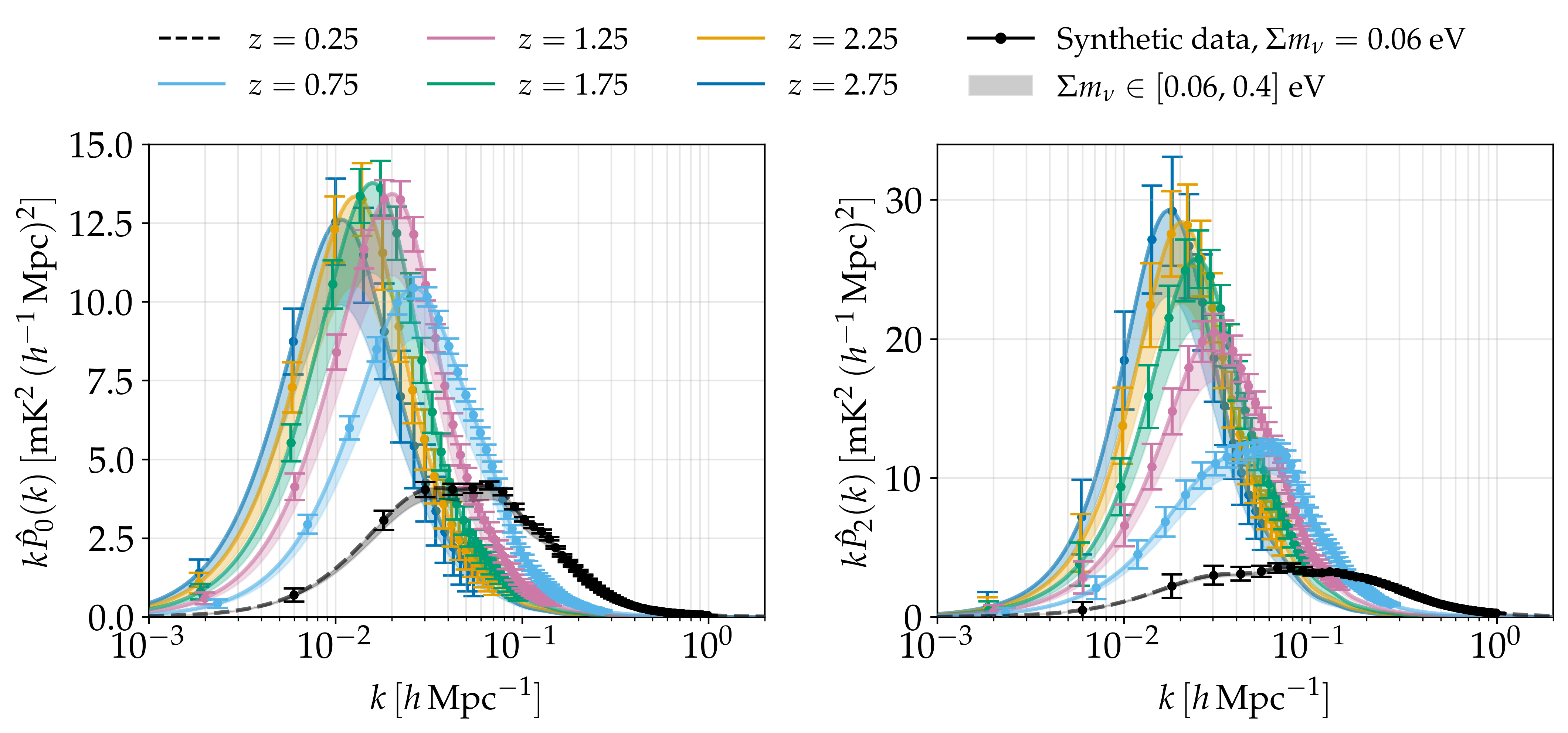}
    \end{subfigure}
    \caption{Synthetic data sets for 21cm power spectrum monopole and quadrupole (\textit{left} and \textit{right} panel, respectively). Shaded regions represent variations in the power spectra for different values of the sum of neutrino masses. We refer to \autoref{sec:21cm_synt_auto} for a complete discussion on how we compute the data sets.}
    \label{fig:data_plots_auto}
\end{figure}
In this work, we construct three data sets to forecast an HI IM observation with the SKAO, as discussed above. For all three data sets, we adopt a Planck 2018 fiducial cosmology \citep{planck:2018}, with three different values for the sum of neutrino masses, $\Sigma m_\nu = 0.06,\; 0.1,\; 0.4\;$eV. We consider the case of $1$ massive neutrino with normal hierarchy. To obtain the fiducial values of the parameters, we run an MCMC analysis with Planck 2018 likelihoods fixing the sum of the neutrino masses at the three values referenced above (see \autoref{sec:likelihood} for further details on the likelihoods). The resulting fiducial values for the cosmological parameter for the three data sets are reported in \autoref{tab:fiducial_cosmos}. The data sets are shown in \autoref{fig:data_plots_auto}. We do not show plots of the signal-to-noise ratios (S/N) of the data sets, as they do not differ significantly from the S/N shown in \citep{Berti:2022ilk}. Therefore, we refer the interested reader to Figure 7 of the above reference.

\subsubsection{21cm and galaxy clustering cross-correlation power spectrum}
\label{sec:modeling cross}
Similarly to what was discussed above for the 21cm multipoles observations, to construct the synthetic cross-correlation data we focus on an SKA-like 21cm IM signal, following again \citep{Bacon:2018}. In particular, we consider the Wide Band 1 Survey described above.

\begin{figure}
    \centering
    \begin{subfigure}{1.0\textwidth}
        \includegraphics[width=1.0\textwidth]{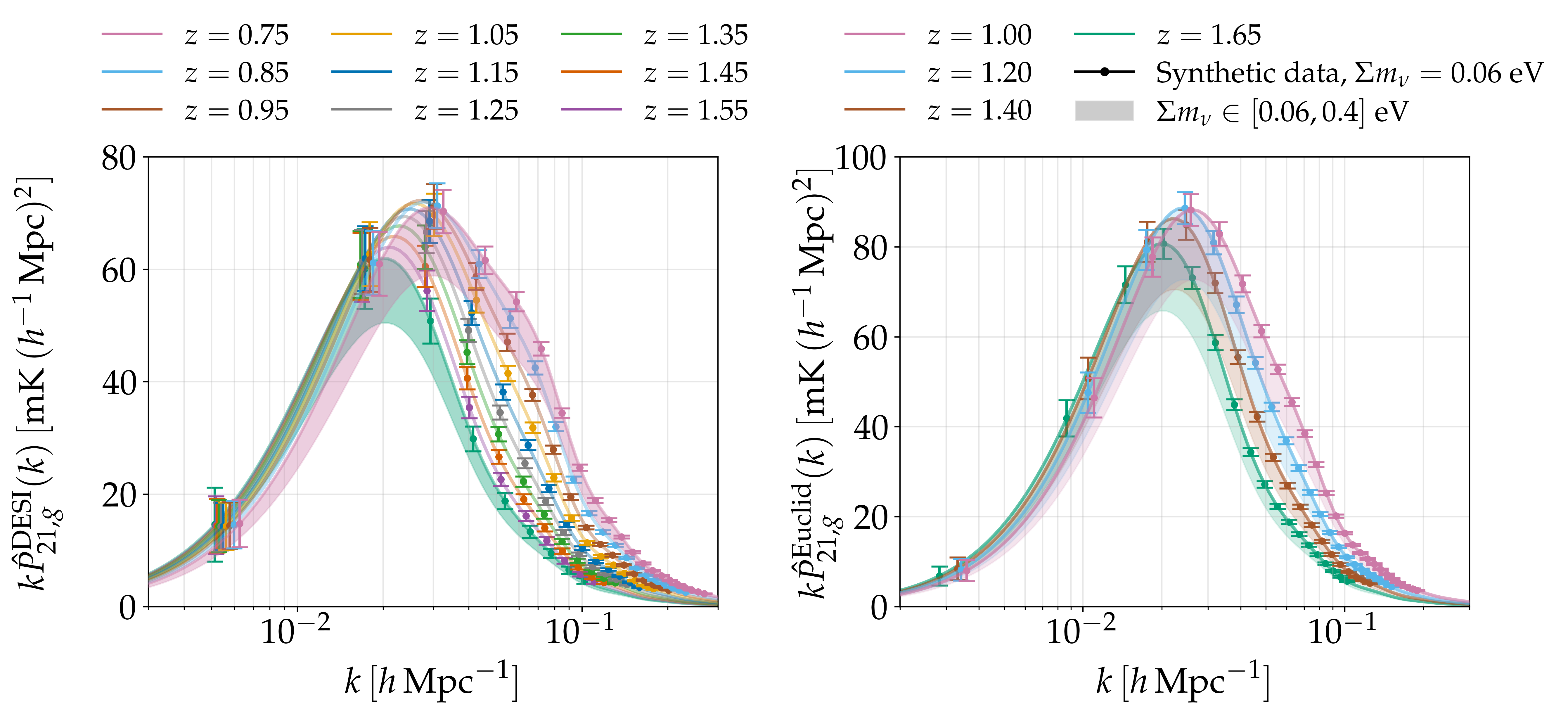}
    \end{subfigure}
    \caption{Synthetic data sets for SKAO$\times$DESI (\textit{left panel}), SKAO$\times$Euclid (\textit{right panel}). Shaded regions for the first and final redshift bins represent variations in the power spectra for different values of the sum of neutrino masses. We refer to \autoref{sec:modeling cross} for a complete discussion on how we compute the data sets.}
    \label{fig:cross_data_plots}
\end{figure}

For Euclid, following \citep{Euclid:2019clj}, we consider observations across four distinct redshift bins within the range $0.9 - 1.8$. We assume an overlapping sky area of $10000 \; \mathrm{deg}^2$ between the SKAO and a Euclid-like survey, which is larger than the overlap between the SKAO and a DESI-like survey, as the SKAO and DESI telescopes are situated in different hemispheres. Hereafter, we will refer simply to the Euclid and DESI surveys, with the understanding that these correspond to the survey configurations described above.
\begin{table} 
    \centering
    \begin{tabular}{l@{\hspace{2cm}}c}
        \toprule
        \multicolumn{2}{c}{SKAO Medium-Deep Band 2} \\ 
        \midrule
        Band frequency range & $0.95\; -\; 1.75\;$ GHz\\
        Corresponding redshift range & $0\;-\; 0.5$\\
        Dish diameter $D_{\mathrm{dish}}$ & $15\;$ [m]\\
        \toprule
        \multicolumn{2}{c}{SKAO Wide Band 1} \\ 
        \midrule
        Band frequency range & $0.35\; -\; 1.05\;$ GHz\\
        Corresponding redshift range & $0.35\;-\; 3$\\
        Dish diameter $D_{\mathrm{dish}}$ & $15\;$ [m]\\
        \toprule
        \multicolumn{2}{c}{SKAO$\times$DESI} \\ 
        \midrule
        Observed redshift range & $0.7\;-\; 1.7$\\
        Overlapping survey area & $5000\; [\mathrm{deg}^2]$\\
        Corresponding $\Omega_{\mathrm{sur}}$ & $1.5\; [\mathrm{sr}]$\\
        \toprule
        \multicolumn{2}{c}{SKAO$\times$Euclid} \\ 
        \midrule
        Observed redshift range & $0.9\;-\; 1.8$\\
        Overlapping survey area & $10000\; [\mathrm{deg}^2]$\\
        Corresponding $\Omega_{\mathrm{sur}}$ & $3.0\; [\mathrm{sr}]$\\
   \bottomrule
    \end{tabular}
    \caption{Survey specifications for SKA-Mid Medium-Deep Band 2 and Wide Band 1 (\citep{Bacon:2018}, DESI ELG (\citep{DESI:2016fyo, Casas:2022vik} and Euclid-like \citep{Euclid:2019clj} surveys. We refer to SKA-Mid as SKAO, to DESI ELG-like as DESI, and Euclid-like as Euclid, for simplicity. }
    \label{tab:surveys_specs}
\end{table}
For both surveys and for each redshift bin, we compute $k_{\mathrm{min}}$ and $k_{\mathrm{max}}$ as described previously. The error on the cross-correlation power spectrum at each point is estimated as \citep{Cunnington:2022uzo, Smith:2008ut}
\begin{equation}
    \sigma_{21,\mathrm{g}}(z, k) = \frac{1}{\sqrt{2N_{\mathrm{modes}}(z, k)}}\, \sqrt{\hat{P}^2_{21,\mathrm{g}}(z, k) + \Bigl(\hat{P}_{21}(z, k) + P_{\mathrm{N}}(z)\Bigr)\Bigl(\hat{P}_{\mathrm{g}}(z, k) + \frac{1}{\overline{n}_\mathrm{g}}(z)\Bigr)}\,,
    \label{sigma_cross}
\end{equation}
where $\hat{P}_{\mathrm{g}}(z,k,\mu)$ is the observed galaxy power spectrum defined as 
\begin{equation}
    \hat{P}_{\mathrm{g}}(z,k,\mu) = \frac{1}{\alpha^2_{\perp}\alpha_{\parallel}}\,P_{\mathrm{g}}(z,q,\nu),
    \label{obs_galaxy_ps}
\end{equation}
where the galaxy power spectrum $P_{\mathrm{g}}$ is defined as
\begin{equation}
    P_{\mathrm{g}}(z,k,\mu) = \Bigl(b_{\mathrm{g}}(z) + f_{\mathrm{CDM + b}}(k, z)\mu^2\Bigr)^2\, P_{\mathrm{CDM + b}}(z,k,\mu),
\end{equation}
where all quantities have been defined above. We note that in Equation \eqref{sigma_cross} we have included the contribution of the 21cm IM instrumental noise $P_\mathrm{N}$ and the contribution of the galaxy shot noise term, $1/\overline{n}_\mathrm{g}(z)$, where $\overline{n}_\mathrm{g}(z)$ is the galaxy number density. As we did for the SKA-like survey, we construct three data sets for both the DESI and the Euclid surveys, each generated with the three different fiducial values for the neutrino masses. The data sets are shown in \autoref{fig:cross_data_plots}. As we mentioned above for the 21cm IM synthetic data sets, we do not report the signal-to-noise ratio as it does not differ significantly from what was found in \citep{Berti:2023viz}, where the constraining power on the $\Lambda$CDM model for analogous cross-correlation data sets with a fixed neutrino mass was investigated. As expected, the SKAO$\times$Euclid data sets have a higher S/N as a result of the larger overlapping sky area compared to the SKAO$\times$DESI survey. For more details, we refer the reader to Figure 2 of the above reference.

\subsection{Numerical analysis}
\label{sec:numerical_analysis}
In order to explore the constraining power of future surveys through the synthetic data sets presented in the previous sections, we build a Gaussian likelihood and perform a Bayesian analysis. The posterior probability distribution for a set of parameters $\Theta$, given a data set $\mathcal{D}$, can be computed as
$\mathcal{P}(\Theta \vert \mathcal{D}) \propto \mathcal{L}(\mathcal{D}\vert \Theta)\, p(\Theta)$, where $\mathcal{L}(\mathcal{D}\vert \Theta)$ is the likelihood and $p(\Theta)$ is the prior probability distribution of the parameters. The resulting posterior distribution is a high-dimensional function that can be sampled using MCMC techniques. 

\subsubsection{Likelihood function implementation}
\label{sec:likelihood}
Given $\Theta(z) = \bigl(\hat{P}_X (z,k_1), \ldots ,\hat{P}_X (z,k_N) \bigr)$, a set of measurements at scales $\{k_1, \ldots, k_N\}$ and at a corresponding redshift $z$, the logarithmic likelihood (or log-likelihood) is computed as 
\begin{equation}
    -\ln{[\mathcal{L}]} = \frac{1}{2}\sum_z\, \Delta\Theta(z)^{\mathrm{T}} C^{-1}(z) \Delta\Theta(z),
    \label{likelihood}
\end{equation}
where $\Delta\Theta(z) = \Theta^{\mathrm{th}}(z) - \Theta^{\mathrm{obs}}(z)$ is the difference between the values of the observables predicted by the theory and the observed ones, $C(z)$ is the covariance matrix and we sum over the non-overlapping redshift bins which we consider to be uncorrelated. {Note that the covariance matrix is analytically estimated following \cite{Bernal:2019jdo}, for each fiducial cosmology. While in our forecast analysis of simulated data sets with known cosmology, we assume the covariance matrix to be fixed, in the analysis with real data it will be estimated, possibly through an iterative approach. We anticipate that an additional systematic error will be introduced in the analysis, due to the small differences between the fiducial (used to estimate the covariance matrix) and true cosmologies. However, we expect this to be a subdominant effect, as the covariance matrix is expected to shift only by sub-percent ($\lesssim 1\%$)\footnote{We estimate this tentative systematic uncertainty by shifting the cosmology through $\sigma_{\Omega_m}/\Omega_m \sim 1\%$ (as constrained with a fixed covariance matrix, see \cref{tab:constraints_P21}) and assessing that the maximum variation in the diagonal elements of covariance matrix to be well within sub-percent level. } level. } 

To perform the MCMC analysis, we use the likelihood code developed in \citep{Berti:2021ccw,Berti:2022ilk, Berti:2023viz} and update it to be integrated in the MCMC sampler \texttt{Cobaya}\footnote{See \url{https://cobaya.readthedocs.io/en/latest/index.html}.}\citep{Torrado:2020dgo}. Our likelihood code, which we named \texttt{topk}, will be made publicly available upon publication. In the MCMC analysis, we vary the six cosmological parameters that describe the $\Lambda$CDM model, $\{\Omega_b h^2$, $\Omega_c h^2$, $ n_\mathrm{s}$, $\ln{(10^{10} A_s)}$, $\tau$, $100\,\theta_{\mathrm{MC}}\}$, as well as the sum of the neutrino masses $\Sigma m_\nu$, and for the 21cm power spectrum, the amplitude of the shot-noise power spectrum in each redshift bin, $P_{\mathrm{SN},i}$. For all sampled parameters, we assume sufficiently wide flat priors, as shown in \autoref{tab:priors}. Furthermore, following previous works \citep{Berti:2021ccw, Berti:2022ilk, Berti:2023viz}, we incorporate nuisance parameters into the analysis. Specifically, some of the parameters that appear in Equation \eqref{21cm_ps} and Equation \eqref{cross_PS}, such as $T_\mathrm{b}$, $b_{\mathrm{HI}}$ and $b_{\mathrm{g}}$, may need to be treated as unconstrained quantities in a pessimistic scenario in which their values are not known. To account for this, we allow for combinations of these parameters to vary in the MCMC sampling and then, in the final results, we marginalize over their contributions. For the 21cm power spectrum, we consider as nuisances the two combinations $\overline{T}_\mathrm{b}b_{\mathrm{HI}} \sigma_8 (z)$ and $\overline{T}_\mathrm{b} f \sigma_8 (z)$. Similarly, for the cross-correlation power spectrum we consider the three combinations $\sqrt{\overline{T}_\mathrm{b}} b_{\mathrm{HI}} \sigma_8 (z)$, $\sqrt{\overline{T}_\mathrm{b}}f \sigma_8 (z)$ and $\sqrt{\overline{T}_\mathrm{b}}b_{\mathrm{g}} \sigma_8 (z)$. We note that in order to rewrite the 21cm and cross-correlation power spectra in terms of these parameters combination, we renormalize the matter power spectrum as $P_m/\sigma_8^2$. The most general approach when dealing with nuisance parameters is to avoid assuming any redshift evolution, instead sampling the value of each parameter independently within each redshift bin. However, this approach introduces a large number of additional parameters to sample, which substantially increases the computational time required for the MCMC procedure to converge. An alternative approach that addresses this issue is to assume a parametrization for the redshift evolution of the nuisances. In our analysis, we use a $3^{\mathrm{rd}}$-degree polynomial to parametrize the redshift evolution of the nuisances that enter the 21cm power spectrum
\begin{equation}
    \overline{T}_\mathrm{b}b_{\mathrm{HI}} \sigma_8 (z),\; \overline{T}_\mathrm{b} f \sigma_8 (z) = az^3 + bz^2 + cz + d,
    \label{cubic_pol_param}
\end{equation}
and a $2^{\mathrm{nd}}$-degree polynomial to parametrize the evolution of the nuisance parameters that enter the cross-correlation power spectrum, that is
\begin{equation}
    \sqrt{\overline{T}_\mathrm{b}}b_{\mathrm{HI}} \sigma_8 (z), \; \sqrt{\overline{T}_\mathrm{b}}f \sigma_8 (z),\; \sqrt{\overline{T}_\mathrm{b}}b_{\mathrm{g}} \sigma_8 (z) = az^2 + bz + c.
    \label{quad_pol_param}
\end{equation}
Thus, in our analysis, we treat the coefficients of the polynomials as nuisance parameters in order to reduce the number of sampled parameters. This approach is particularly beneficial for the IM and the cross-correlation analysis of the SKA$\times$DESI survey. For the IM, where we have $6$ redshift bins, introducing the redshift parametrization of the nuisance parameters reduces their number from $12$ to $8$. The advantage of this approach is even more pronounced for the SKA$\times$DESI cross-correlation, which involves $10$ redshift bins; in this case, the parametrization reduces the number of nuisance parameters from $30$ to $9$. For SKA$\times$Euclid cross-correlation, with only $4$ redshift bins, the reduction is less significant, from $12$ to $9$. Nevertheless, we apply the same methodology across all cases to ensure consistency and to facilitate more realistic direct comparisons. 

\begin{table}
	\centering
	\begin{tabular}{ll}
 \toprule
 Parameters & Prior\\ 
  \midrule
   $\Omega_b h^2$ & $\mathcal{U}[0.005, 0.1]$ \\
   $\Omega_c h^2$ & $\mathcal{U}[0.001, 0.99]$ \\
   $n_\mathrm{s}$ & $\mathcal{U}[0.8, 1.2]$ \\
   $\ln (10^{10} A_s)$ & $\mathcal{U}[1.61, 3.91]$ \\
   $\tau$ & $\mathcal{U}[0.01, 0.8]$ \\
   $100\,\theta_\mathrm{MC}$ & $\mathcal{U}[0.5, 10]$ \\
   $\Sigma m_\mathrm{\nu}$ (eV) & $\mathcal{U}[0.059, 1]$\\ 
   \bottomrule
	\end{tabular}
    
 	\caption{Cosmological parameters and priors used in the analysis. All the priors are uniform in the given range.}
    \label{tab:priors}
\end{table}

\subsubsection{External likelihoods and data sets}
\label{sec:data_sets}
We combine our synthetic IM and cross-correlation data sets with Planck 2018 observations \citep{planck:2018, planck:2018like}\footnote{Note that the more recent CMB likelihood, \texttt{HiLLiPoP}, along with the final Planck release PR4~\cite{Tristram:2023haj} provides mildly tighter constraints on the cosmological parameters. However, we stick with the PR3 (Planck 2018) likelihood for ease of comparison with earlier works. Note also that, interestingly, the more recent PR4 with \texttt{HiLLiPoP} likelihood provides mildly looser constraints on the sum of neutrinos masses $\Sigma m_{\nu} < 0.39 $ eV at $95\%$ C.L. \cite{Tristram:2023haj}, in comparison to the earlier Planck 2018   constraint of $\Sigma m_{\nu} < 0.26 $ eV \cite{planck:2018}. This also justifies our choice of higher mass $\Sigma m_{\nu} = 0.4$ forecast analysis, which is the $95\%$ C.L. limit of the PR4 constraints.}. The CMB likelihoods that we use include high-$\ell$ TTTEEE foreground marginalized \texttt{plik}-lite likelihood for multipoles $30 \leq \ell \leq 2508$ for TT and $30 \leq \ell \leq 1996$ for TE and EE. For the low-$\ell$ TT power spectrum, we use the data from the \texttt{Commander} likelihood in the range $2\leq \ell \leq 29$. We also include the low EE polarization power spectrum, referred to as lowE, calculated from the likelihood \texttt{SimAll} in the range $2\leq \ell \leq 29$ and the CMB lensing likelihood. In the remainder of this work, with the label "Planck 2018" we refer to the combination of TTTEEE + low-$\ell$ + lowE + lensing. 

We emphasize that, when the synthetic data are combined with real observations, we are primarily interested in the resulting error bars, which accurately predict the precision with which these parameters can be constrained. The resulting marginalized means are largely uninformative and serve primarily as a check that the method successfully recovers the input fiducial values.

\section{Results}
\label{sec:results}
In this section, we present the results of this analysis. In the following, upper bounds on $\Sigma m_\mathrm{\nu}$ refer to the $95\%$ confidence limits, while full constraints refer to $68\%$ confidence intervals. Throughout all the discussion, we report results obtained by fixing nuisance parameters and we report in parentheses the constraints obtained when nuisances are marginalized upon. We comment on the constraints obtained for the standard cosmological parameters in \autoref{sec:Gen_cons}. We explore the constraining power of the synthetic 21cm IM data sets described above alone and combined with Planck 2018 CMB data in \autoref{sec:21cm results}. In \autoref{sec:cross_results} we present the results of the analysis with the cross-correlation synthetic data sets with and without the addition of CMB data.\\

\subsection{Constraints on the $\Lambda$CDM cosmological parameters}
\label{sec:Gen_cons}
We begin by commenting on the constraints obtained for the standard cosmological parameters, before proceeding to the constraints on the neutrino mass. In \autoref{tab:constraints_P21}, we summarize the constraints obtained utilizing only the 21cm IM data sets, as well as the constraints obtained when combining the 21cm IM data with Planck 2018 CMB data. The results are shown for the three different fiducial values of $\Sigma m_\nu$ considered in our analysis. We find that the constraints on the standard cosmological parameters are significantly improved utilizing only the 21cm IM data and are mildly improved in the joint analysis with CMB, compared to CMB data alone. In particular, we find that the constraints on $H_0$ are improved by an order of magnitude with respect to the CMB-only constraints in \cite{planck:2018}, for all three fiducial cases. Similarly, the total matter density $\Omega_m$ is improved by a factor of $\sim 2$, alongside mild improvements in the $\sigma_8$ constraints. 

This improvement in the the constraints of the background parameters $\{H_0, \Omega_m\}$ is partially due to the fact that the 21cm IM data sets are sensitive to the growth of structure at non-linear scales, which is not well probed by CMB data and mostly aided by the BAO information that is modeled through the AP effect in the 21cm IM data sets. The latter in particular is responsible for the improvement in the constraints on $H_0$ and $\Omega_m$ even when Planck 2018 CMB data is not included. We note that the constraints on $\sigma_8$ are not significantly improved ($\sim 30\%$) unless aided by CMB data, reaching a $\sim 60\%$ smaller uncertainty. While the inclusion of the nuisance parameters (column 4 of \cref{tab:constraints_P21}) mildly increases the uncertainties, they are still significantly stringent than the CMB-only constraints. 

Although we do not perform a dedicated assessment of the constraints on the BAO scales, we briefly discuss only the constraints on them. We find that the BAO scales modeled intrinsically in our formalism through the AP effect and the scales accessed by 21cm observations, in the six redshift bins of the 21cm IM auto-power spectrum synthetic data, are able to provide a detection of the BAO observables with a very high precision. For instance, we find an error on the angular diameter distance $D_A$ ranging from $\sigma_{D_A(z = 0.25)} \sim 0.1\%$ to $\sigma_{D_A(z = 2.75)} \sim 0.3\%$. Similarly, for the Hubble parameter, we find $\sigma_{H(z = 0.25)} \sim 0.1\%$ to $\sigma_{H(z = 2.75)} \sim 0.5\%$. The precisions at $z\sim 2.5$ are comparable to those presented in \cite{Obuljen:2017}, at a similar redshift range. This, in turn, provides us with extremely good constraints on the background $\{H_0, \Omega_{\rm m}\}$ parameters. We note that, in previous works \citep{Berti:2022ilk, Berti:2023viz}, the AP effect was not included in the analysis. However, \citep{Berti:2023viz} explored the impact of adding the AP effect on the constraints on cosmological parameters, finding that the addition of the AP effect leads to tighter constraints. Notably, the improvement that we find in our results compared to those of \citep{Berti:2022ilk, Berti:2023viz} is compatible with the expected improvement obtained by adding the AP effect to the analysis and therefore can be ascribed to that. 

Now turning to the cross-correlation analysis, wherein the SKAO IM data is combined with the DESI and Euclid galaxy surveys, we find that the final constraints as reported in \autoref{tab:constraints_cross_desi} and \autoref{tab:constraints_cross_euclid} are typically of the order of the current Planck-based CMB constraints. These constraints are then weakened by a factor of $\sim 2-3$ when the nuisance parameters are left free to vary. Cross-correlating with the DESI and Euclid galaxy surveys in similar redshift ranges of $z \in 0.7 - 1.7$ and $z \in 0.9 - 1.8$, respectively, we find the constraining power of both the observables to be similar. This is true despite the larger overlapping area of the SKAO$\times$Euclid survey compared to the SKAO$\times$DESI one.

However, for SKAO$\times$DESI we forecast a finer redshift binning that includes effective redshifts lower than for the SKAO$\times$Euclid survey. As the smallest accessible scale in a given redshift bin is set by the effect of the beam smoothing, which is smaller at lower redshifts, the SKAO$\times$DESI survey can probe smaller scales compared to those accessible by the SKAO$\times$Euclid survey at lower redshifts. We find that the cross-correlation analysis with DESI and Euclid-like surveys provides similar constraints on the background parameters as well as on $\Sigma m_\nu$, which are elaborated upon in the following sections.

\subsection{Constraints on $\Sigma m_\nu$ from 21cm auto-power spectrum observations}
\label{sec:21cm results}
\begin{figure}
    \centering
    \includegraphics[width=1.0\textwidth]{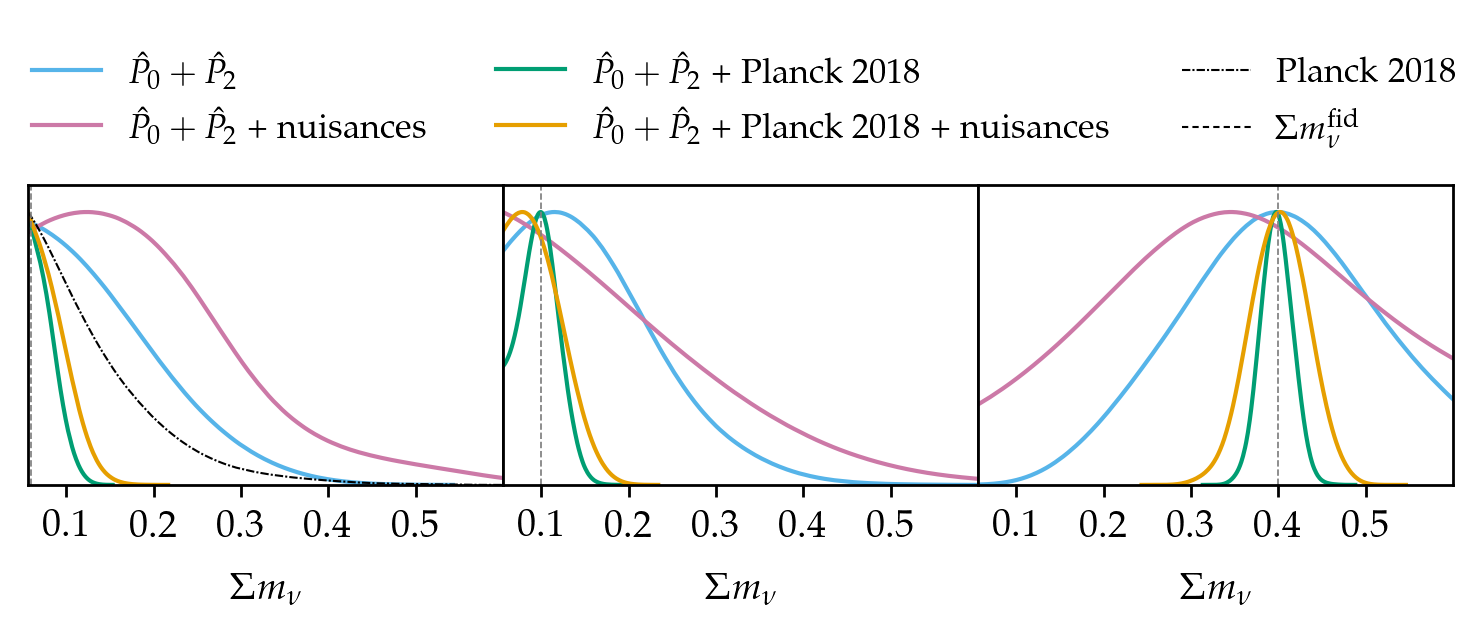}
    \caption{Marginalized 1D posterior distributions of $\Sigma m_\nu$ for the 21cm IM synthetic data sets alone, and combined with Planck 2018 observations. Each panel corresponds to one of the three cosmologies characterized by different fiducial values of the neutrino mass marked by vertical dashed lines, i.e. $\Sigma m_\nu^{\mathrm{fid}} = 0.06$ eV (\textit{left panel}), $\Sigma m_\nu^{\mathrm{fid}} = 0.1$ eV (\textit{middle panel}) and $\Sigma m_\nu^{\mathrm{fid}} = 0.4$ eV (\textit{right panel}).}
    
    \label{fig:SKA_results}
\end{figure}

 We show the forecasted marginalized 1D posterior distributions for the sum of neutrino masses $\Sigma m_\nu$ obtained from the 21cm IM synthetic data sets in \autoref{fig:SKA_results}. In the normal hierarchy scenario, a minimum mass for neutrinos of $\Sigma m_\nu = 0.059$ eV is required, therefore, in our analysis, we always set this value as a lower limit for the prior on the sum of neutrino masses.
 
For the case of $\Sigma m_\nu ^{\mathrm{fid}} = 0.06$ eV, we find an upper limit of $\Sigma m_\nu < 0.287\; (0.425)\: \mathrm{eV}$. The upper limit obtained with nuisance parameters fixed is comparable with the upper limit found by Planck 2018 CMB data of $\Sigma m_\nu < 0.285$ eV. Letting the nuisance parameters free to vary, instead, loosens the constraint. Similarly, for the case of $\Sigma m_\nu ^{\mathrm{fid}} = 0.1$ eV we obtain $\Sigma m_\nu < 0.317\; (0.452)$ eV. Again, the result for nuisances fixed is comparable with the upper limit obtained by Planck 2018. For the case of $\Sigma m_\nu ^{\mathrm{fid}} = 0.4$ eV, we find a full constraint, i.e. $\Sigma m_\nu = 0.41^{+0.11}_{-0.14}\; (0.34^{+0.16}_{-0.14})$ eV. In this case, a comparison with the result of Planck 2018 does not make sense, as $\Sigma m_\mathrm{\nu}=0.4$ eV is excluded by Planck data.
\begin{figure}
    \centering
    \includegraphics[width=1.0\textwidth]{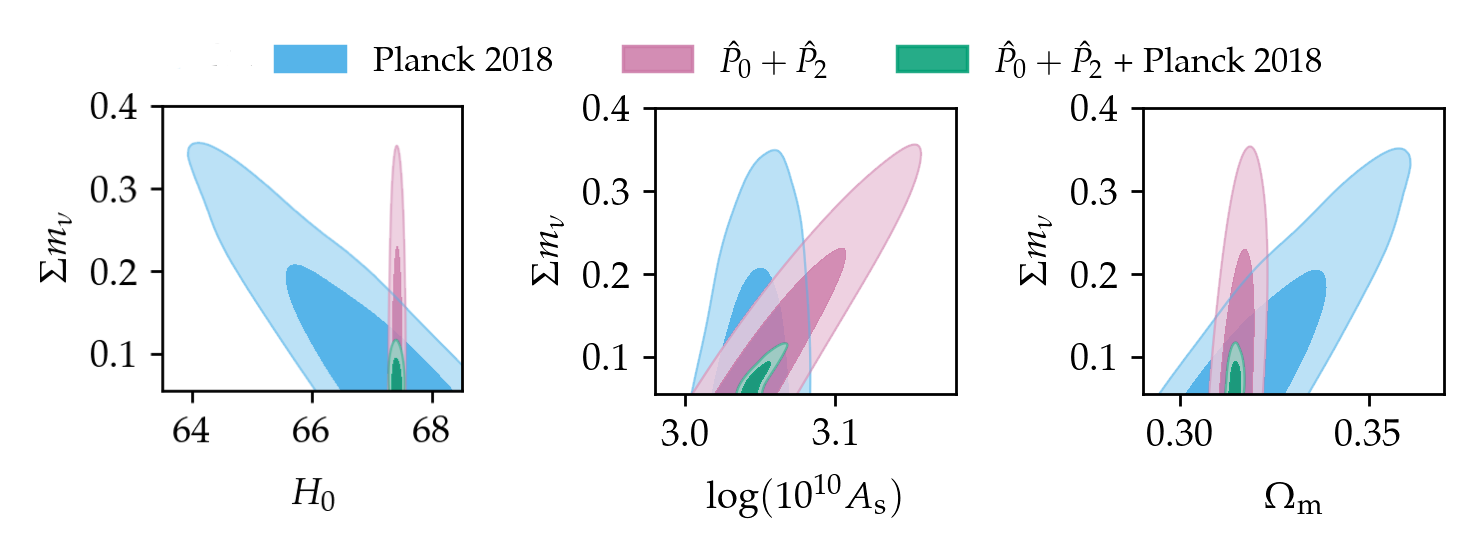}
    \caption{Marginalized 2D posterior distributions in the $\Sigma m_\nu$--$H_0$, $\Sigma m_\nu$--$\Omega_\mathrm{m}$ and $\Sigma m_\nu$--$\sigma_8$ planes for Planck 2018 data and the 21cm IM synthetic data sets alone, and combined. We show the case of $\Sigma m_\nu^{\mathrm{fid}} = 0.06$ eV.}
    \label{fig:SKA_2D}
\end{figure}

Additionally, we include Planck CMB data in our analysis, as previously described in \autoref{sec:data_sets}, and we report the obtained forecasted posteriors in \autoref{fig:SKA_results}. The constraints obtained on $\Sigma m_\mathrm{\nu}$ by including also CMB data are significantly tighter than what we found for the synthetic data sets alone. In addition to the reasons discussed previously in \autoref{sec:Gen_cons}, the improvement found by adding CMB data to our synthetic 21cm IM data sets is due to the fact that for Planck CMB data, $\Sigma m_\mathrm{\nu}$ shows a strong anti-correlation with $H_0$. We show in \autoref{fig:SKA_2D} the marginalized 2D contours in the $\Sigma m_\mathrm{\nu}$-$H_0$, $\Sigma m_\mathrm{\nu}$- $\Omega_\mathrm{m}$ and $\Sigma m_\mathrm{\nu}$-$\sigma_8$ planes. As we can see in the first panel, the 21cm IM data provide a very stringent constraint on $H_0$ which, when combining with CMB data, improves the result obtained on the sum of neutrino masses and removes the degeneracy in the $H_0$-$\Sigma m_\mathrm{\nu}$ plane exhibited by CMB data. In particular, for the case with $\Sigma m_\nu ^{\mathrm{fid}} = 0.06$ eV, we find $\Sigma m_\nu < 0.105 \; (0.126)$, which represents improvements of factors $\sim 2.7$ and $\sim 3.4$ for the case of nuisances fixed and left free, respectively. For $\Sigma m_\nu ^{\mathrm{fid}} = 0.1$ eV we find that, for fixed nuisances, we obtain $\Sigma m_\nu = 0.098\pm 0.022$ eV. When nuisances are left free to vary, we find the upper limit $\Sigma m_\nu < 0.151$ eV, marking an improvement of a factor $\sim 3$. Finally, in the case of $\Sigma m_\nu ^{\mathrm{fid}} = 0.4$ eV, we find $\Sigma m_\nu = 0.398\pm0.018 \; (0.401\pm0.034)$ eV, which represents an improvement in the constraint of factors of  $\sim 6.1$ and $4.1$, respectively.

We conclude that by combining 21cm IM data with CMB data, we are able to break degeneracies between $\Sigma m_\mathrm{\nu}$ and other parameters, such as $H_0$, improving the constraints on the sum of neutrino masses by a factor that ranges from $\sim 3$ to $\sim 6$, depending on the fiducial value of $\Sigma m_\mathrm{\nu}$.

A full list of constraints on all cosmological parameters obtained with the 21cm IM synthetic data sets is reported in \autoref{tab:constraints_P21}. 

\subsection{Constraints on $\Sigma m_\nu$ from the 21cm and galaxy clustering cross-correlation observations}
\label{sec:cross_results}
\begin{figure}
    \centering
    \begin{subfigure}{1.0\textwidth}
        \includegraphics[width=1.0\textwidth]{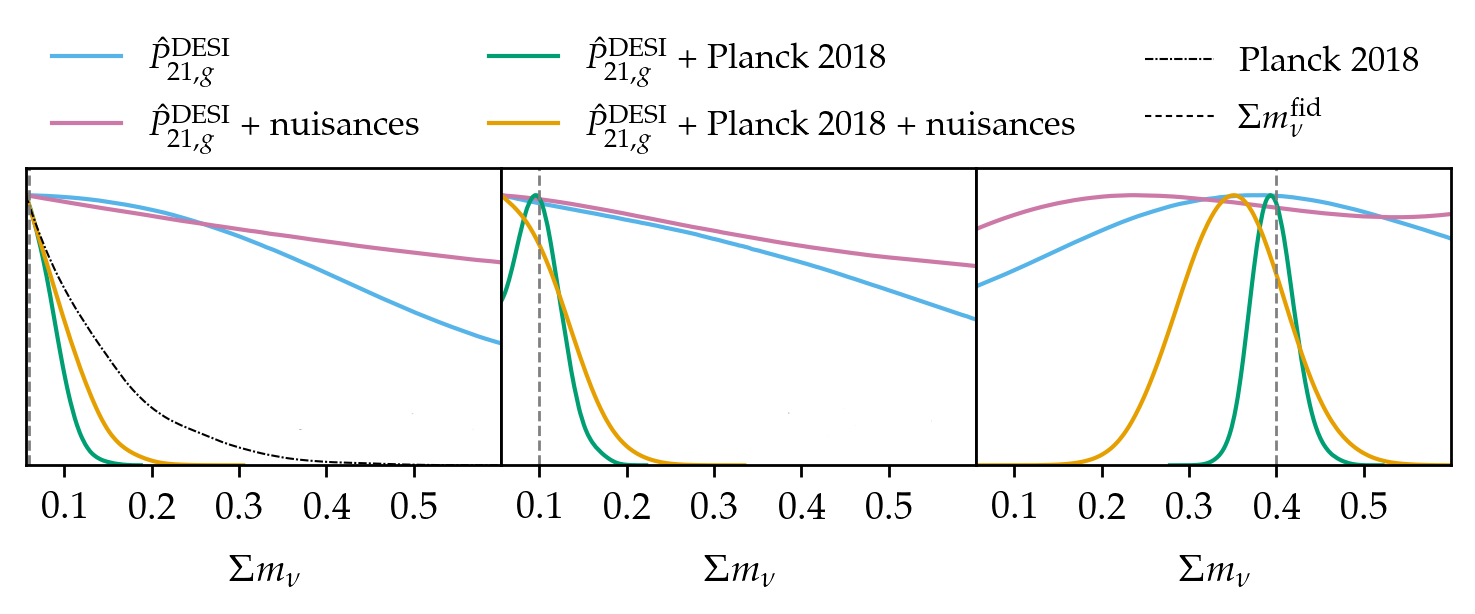}
    \end{subfigure}
    \vspace{0.2cm} 
    \begin{subfigure}{1.0\textwidth}
        \includegraphics[width=1.0\textwidth]{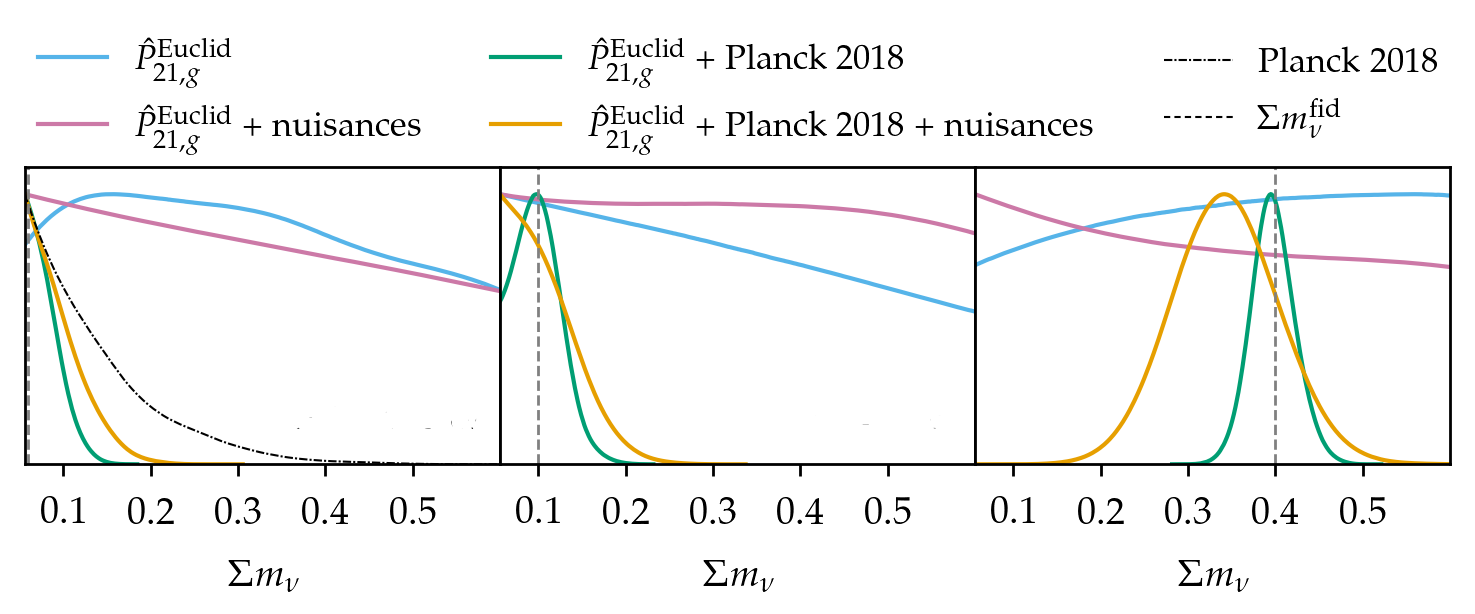}
    \end{subfigure}
    \caption{Marginalized 1D posterior distributions of $\Sigma m_\nu$ for the SKA$\times$DESI survey (\textit{top row}) and SKA$\times$Euclid survey (\textit{bottom row}). Each panel corresponds to one of the three cosmologies characterized by different fiducial values of the neutrino mass marked by vertical dashed lines, i.e. $\Sigma m_\nu^{\mathrm{fid}} = 0.06$ eV (\textit{left panel}), $\Sigma m_\nu^{\mathrm{fid}} = 0.1$ eV (\textit{middle panel}) and $\Sigma m_\nu^{\mathrm{fid}} = 0.4$ eV (\textit{right panel}).}
    \label{fig:cross_results}
\end{figure}
In \autoref{fig:cross_results}, we show the forecasted marginalized 1D posterior distributions of $\Sigma m_\mathrm{\nu}$ obtained from the cross-correlation synthetic data sets. As discussed in \autoref{sec:Gen_cons}, the results obtained for the SKAO$\times$DESI and SKAO$\times$Euclid cross-correlation data sets are very similar. 

We can see that, including only the cross-correlation data sets in the analysis, we have no constraining power on the sum of neutrino masses. This can be explained as a combination of multiple factors: firstly, both the SKAO$\times$Euclid and SKAO$\times$DESI cross-correlation data sets present lower values of the signal-to-noise ratio compared to the 21cm IM observations and therefore have less constraining power. Secondly, the redshift ranges explored by the two cross-correlation data sets are more limited compared to those explored by the auto-power spectrum one. In particular, the cross-correlation data sets do not have access to redshifts lower than $z\simeq 0.75$ and therefore cannot probe the non-linear scales as well as the IM observations do. This results in overall worse constraints on all cosmological parameters. However, we stress that detections of the 21cm IM signal in cross-correlation are expected to be less affected by systematics compared to the auto-correlation detections and therefore will be available in the near future. Notably, the SKA-Mid precursor, MeerKAT, has already reported a detection of the 21cm IM signal in cross-correlation with the WiggleZ galaxy survey \citep{Cunnington:2022uzo, 2010MNRAS.401.1429D, 2018MNRAS.474.4151D}.

Furthermore, adding the CMB data to our analysis drastically improves the constraints obtained. For the case of $\Sigma m_\nu ^{\mathrm{fid}} = 0.06$ eV, we find $\Sigma m_\nu < 0.116\; (0.155)\: \mathrm{eV}$ and $\Sigma m_\nu < 0.117\; (0.156)\: \mathrm{eV}$ for the SKAO$\times$DESI and SKAO$\times$Euclid data sets, respectively. Similarly, for the case of $\Sigma m_\nu ^{\mathrm{fid}} = 0.1$ eV, we get $\Sigma m_\nu = 0.099^{+0.020}_{-0.033}\; (< 0.177)\: \mathrm{eV}$ for the SKAO$\times$DESI data set and $\Sigma m_\nu = 0.100^{+0.021}_{-0.032}\; (< 0.180)\: \mathrm{eV}$ for the SKAO$\times$Euclid data set. Finally, for the case $\Sigma m_\nu ^{\mathrm{fid}} = 0.4$ eV we obtain the following confidence intervals, $\Sigma m_\nu = 0.396^{+0.023}_{-0.026}\; (0.349\pm 0.060)$ and $\Sigma m_\nu = 0.397^{+0.023}_{-0.046}\; (0.343\pm 0.062)$, for the SKAO$\times$DESI and SKAO$\times$Euclid synthetic data sets, respectively.

These results are compatible with the ones obtained with the 21cm IM synthetic data sets, especially in the case of fixed nuisance parameters. Similarly to what we discussed in the previous section, combining our synthetic cross-correlation data sets and CMB data leads to the breaking of degeneracies, drastically improving the constraints. For the sake of brevity, we do not show the contours for the SKAO$\times$Euclid data sets, which are very similar to those of SKAO$\times$DESI. 

In light of these results, we conclude that by combining cross-correlation observations with CMB data we are able to break degeneracies between $\Sigma m_\mathrm{\nu}$ and other parameters, leading to constraints on the sum of neutrino masses that are competitive with the ones obtained from 21cm IM auto-power spectrum syntethic data.

\section{Conclusions}
\label{sec:conclusion}
\begin{table}
	\centering
        \renewcommand{\arraystretch}{1.3}
	\begin{tabular}{lccc} 
	\toprule 
 Likelihoods & $\Sigma m_\nu^{\rm fid} =0.06\,$eV   & $\Sigma m_\nu^{\rm fid} =0.1\,$eV   &  $\Sigma m_\nu^{\rm fid} = 0.4\,$eV\\ 
  \midrule
   $\Hat{P}_0 + \Hat{P}_2$ & $< 0.287$ & $< 0.317$ & $0.41^{+0.11}_{-0.14}$\\
   \ \ \ \  + nuisances &  $< 0.425$ & $< 0.452$  & $0.34^{+0.16}_{-0.14}$\\
   \midrule
   Planck 2018 \\
   \ \ \ \  + $\Hat{P}_0 + \Hat{P}_2$ & $< 0.105$ & $0.098\pm 0.022$ & $0.398\pm 0.018$\\
   \ \ \ \ \ \ \ \ + nuisances &  $< 0.126$ & $ < 0.151$  & $0.401\pm 0.034$\\
   \midrule
   Planck 2018 \\
   \ \ \ \  + $\Hat{P}_{21,\rm g}^{\rm DESI}$ & $< 0.116$ & $0.099^{+0.020}_{-0.033}$ & $ 0.396^{+0.023}_{-0.026}$\\
   \ \ \ \ \ \ \ \ + nuisances & $< 0.155$ & $<0.177$ & $0.349\pm 0.060$\\
   \midrule
   Planck 2018\\
   \ \ \ \  + $\Hat{P}_{21,\rm g}^{\rm Euclid}$ & $< 0.117$ & $0.100^{+0.021}_{-0.032}$ & $0.397^{+0.023}_{-0.026}$\\
   \ \ \ \ \ \ \ \ + nuisances & $< 0.156$ & $<0.180$ & $0.343\pm 0.062$\\
   \bottomrule
	\end{tabular}

 	\caption{Marginalized constraints ($95\%$ confidence level for upper limits, $68\%$ confidence intervals for full constraints) on the sum of neutrino masses, $\Sigma m_\mathrm{\nu}$, for different combinations of the synthetic data sets. Marginalized 1D posteriors are shown in Figure \ref{fig:SKA_results} and in Figure \ref{fig:cross_results}. We do not report constraints from SKAO$\times$DESI and SKAO$\times$Euclid alone since the parameters are left unconstrained in this case. 
    }
    \label{tab:constraints}
\end{table}
In this work, we forecast the constraining power of future 21cm intensity mapping observations with the SKAO on the sum of neutrino masses, $\Sigma m_\mathrm{\nu}$, by
probing the effects of neutrino free-streaming on the matter power spectrum observed through the 21cm intensity mapping and 21cm IM-galaxy clustering cross-correlation power spectra. We focus on the cosmological surveys proposed for the SKA-Mid telescope and also consider
future cross-correlation measurements of the SKAO IM signal with stage-IV galaxy surveys such as DESI and Euclid. We model the 21cm IM and cross-correlation power spectra in cosmologies with massive neutrinos, building upon the formalism presented in \citep{Berti:2021ccw, Berti:2022ilk, Berti:2023viz}. We assess the constraining power of these probes independently and in combination with Planck CMB observations.

Following the SKAO Red Book \citep{Bacon:2018}, we consider single-dish observations with the SKA-Mid telescope. Specifically, we combine a Wide Band 1 survey and a Medium-Deep Band 2 survey in order to construct synthetic 21cm IM observations spanning the redshift range $0 - 3$. We also cross-correlate these observations with spectroscopic Euclid-like \citep{Euclid:2019clj} and DESI-like \citep{Casas:2022vik, DESI:2016fyo} surveys in the redshift range $0.7-1.7$. For each of these three types of data sets, we consider three fiducial values for the sum of neutrino masses: $\Sigma m_\nu = 0.06,\; 0.1,\; 0.4\;$ eV and we construct synthetic data sets for each of them. To explore their constraining power, we implement a likelihood function for the 21cm IM and cross-correlation power spectra, fully integrated in the MCMC sampler \texttt{Cobaya}, that will be made publicly available upon publication. Furthermore, we include nuisance parameters in our analysis to mimic a more realistic scenario in which a perfect knowledge of the involved astrophysics, e.g. the exact values of the brightness temperature and the 21cm bias, is lacking.

The results from our analysis are summarized in~\autoref{tab:constraints}. In the following, we report constraints obtained with nuisance parameters held fixed, and in parentheses, the corresponding bounds obtained when the nuisances are allowed to vary. We investigate the constraining power of the synthetic data sets alone and combined with CMB observations. We find that the 21cm IM synthetic data sets alone are able to provide upper limits on $\Sigma m_\nu$ competitive with CMB observations. 
For a standard fiducial value of the neutrino mass, we find the following constraints $\Sigma m_\nu < 0.287 \; (0.425)$ eV. These bounds are improved in more exotic scenarios with a high neutrino mass, i.e. we find $\Sigma m_{\mathrm{\nu}} = 0.41^{+0.11} _{-0.14}\; (0.34^{+0.16} _{-0.14})$ eV for $\Sigma m_\nu^{\mathrm{fid}} = 0.4$ eV. When CMB data are added, the constraints significantly improve, also with respect to CMB alone results, yielding e.g. $\Sigma m_\nu < 0.105 \;(0.126)$ eV for $\Sigma m_\nu^{\mathrm{fid}} = 0.06$ eV.

Conversely, we find that cross-correlation data sets alone are not able to constrain $\Sigma m_\mathrm{\nu}$. However, when combined with CMB data, the constraints that we get are competitive with those obtained by combining CMB and 21cm IM synthetic data sets. For $\Sigma m_\nu^{\mathrm{fid}} = 0.06$ eV we find $\Sigma m_\nu < 0.116 \; (0.155)$ eV and $\Sigma m_\nu < 0.117\; (0.156)$ eV from combinations of CMB and the SKAO$\times$DESI and SKAO$\times$Euclid cross-correlation data sets, respectively. 

Although the full cosmological information is included in the auto-power spectrum signal, the detection of the 21cm in cross-correlation with galaxy clustering is less affected by systematics effects. For this reasons, cross-correlation measurements are expected to be available sooner, as shown by the recent promising results obtained with the SKAO precursor MeerKAT \citep{Cunnington:2022uzo, MeerKLASS:2024ypg}. Our analysis shows that, when combined with CMB data, a multiple bin detection of the 21cm signal in cross-correlation with stage-IV galaxy clustering surveys will provide already strict constraints on the neutrino mass. This, thus, supports the motivation for cross-correlation studies.

We conclude that the forecasted 21cm IM and 21cm-galaxy clustering cross-correlation measurements, when combined with complementary probes such as Planck CMB data, would be able to provide tight constraints on $\Sigma m_\mathrm{\nu}$. Our findings thus indicate that future 21cm intensity mapping observations, especially when combined with CMB data and other probes of the large-scale structures, will be able to provide competitive constraints on $\Sigma m_\mathrm{\nu}$.

\section*{Data availability}
The likelihood code \texttt{topk} is available at \url{https://github.com/mberti94/topk}. The new version of the code that includes neutrino cosmologies used in the analysis will be made publicly available upon publication.
Access to the synthetic data sets is available upon reasonable request to the corresponding author.

\acknowledgments
The authors would like to thank Gabriele Parimbelli, Steve Cunnington and José Fonseca for useful discussions. GA, BSH and MV are supported by the INFN INDARK grant. MB acknowledges support from the Swiss National Science Foundation. MV is also supported by the Fondazione ICSC, Spoke 3 Astrophysics and Cosmos Observations, National Recovery and Resilience Plan Project ID CN\_00000013 ``Italian Research Center on High-Performance Computing, Big Data and Quantum Computing'' funded by MUR Missione 4 Componente 2 Investimento 1.4: Potenziamento strutture di ricerca e creazione di "campioni nazionali di R\&S (M4C2-19 )" - Next Generation EU (NGEU).

\appendix
\renewcommand{\thetable}{\Alph{section}.\arabic{table}}
\section{Scale dependence of the growth rate $f$}
\begin{figure}
    \centering
    \begin{subfigure}{0.49\textwidth}
        \includegraphics[width=\linewidth]{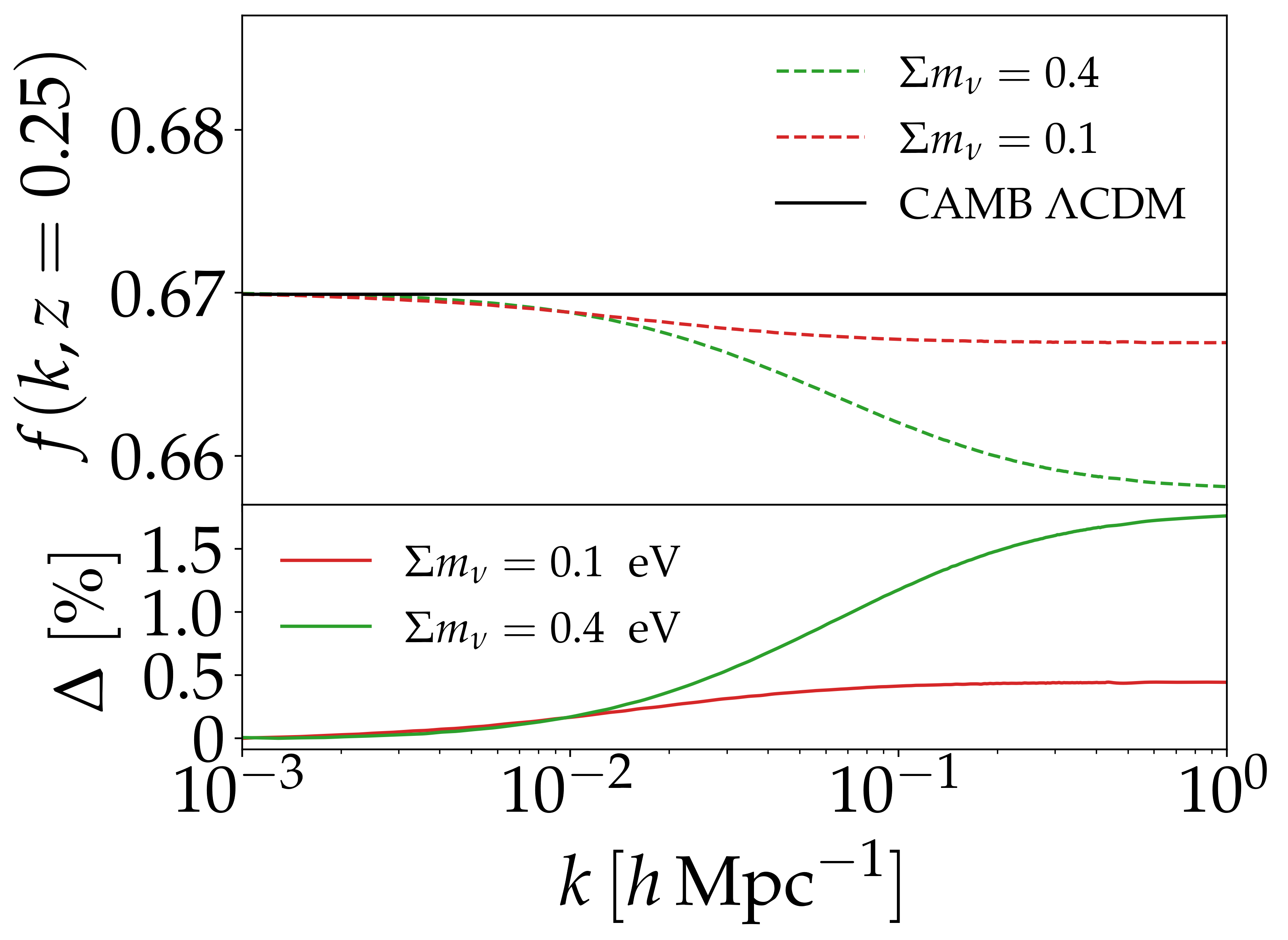}
    \end{subfigure}
    \hfill
    \begin{subfigure}{0.49\textwidth}
         \includegraphics[width=\linewidth]{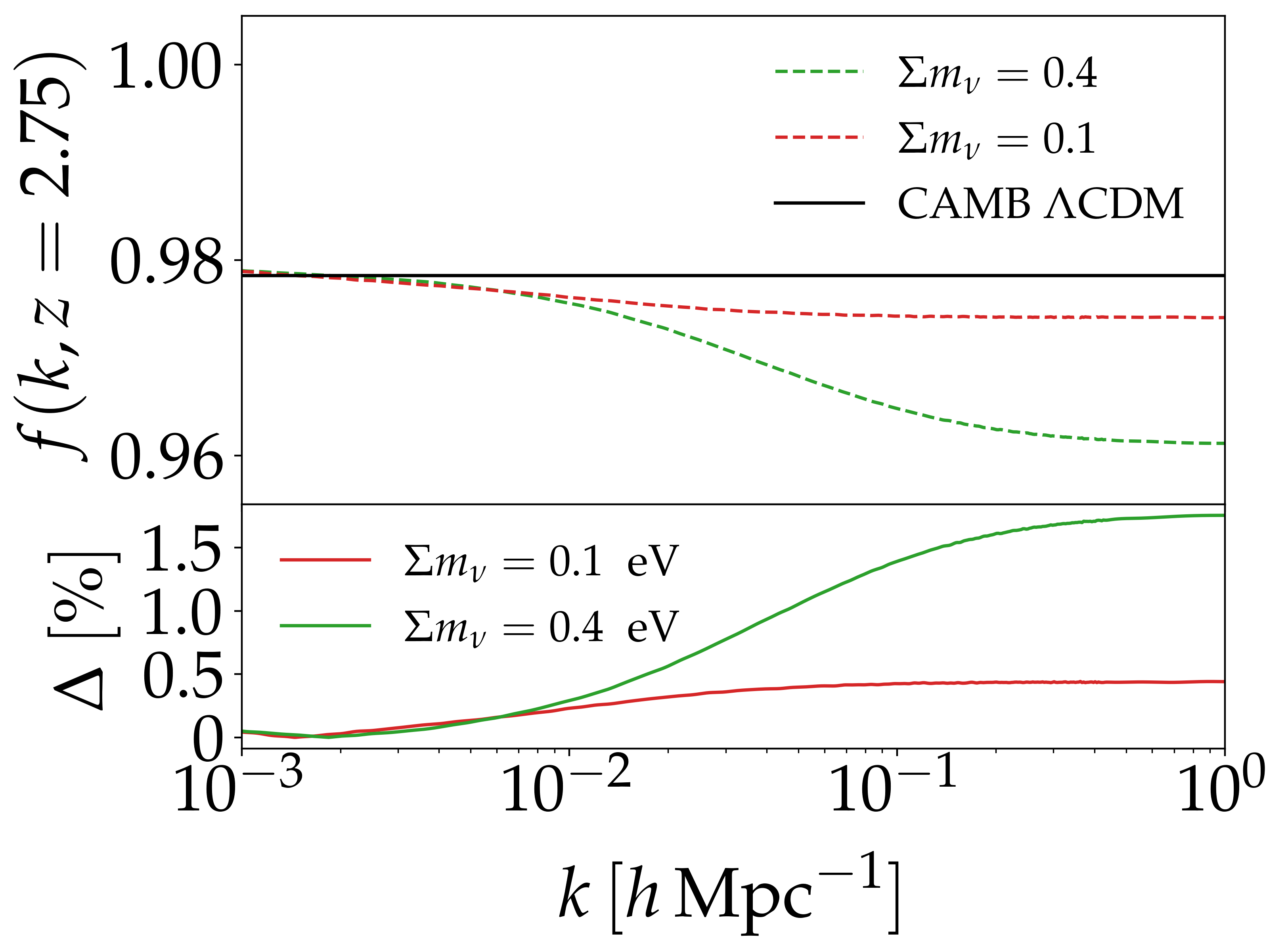}
    \end{subfigure}
    \caption{Scale dependence of the growth rate for different values of the sum of neutrino masses for $z=0.25$ (\textit{left panel}) and $z=2.75$ (\textit{right panel}). The black solid line represents $f(z)$ computed with CAMB for a neutrino-less $\Lambda$CDM model, the colored dashed lines represent the full computations of $f(z,k)$ for a $\Lambda$CDM model with $\Sigma_\mathrm{\nu} = 0.1, \, 0.4$ eV. In the lower panel of the plots, we show the percentage difference of the full calculation with respect to the CAMB one. We fix $\Omega_m$ for all curves shown here.}
    \label{fig:f_cc}
\end{figure}
\begin{figure}
    \centering
    \includegraphics[width=1.0\textwidth]{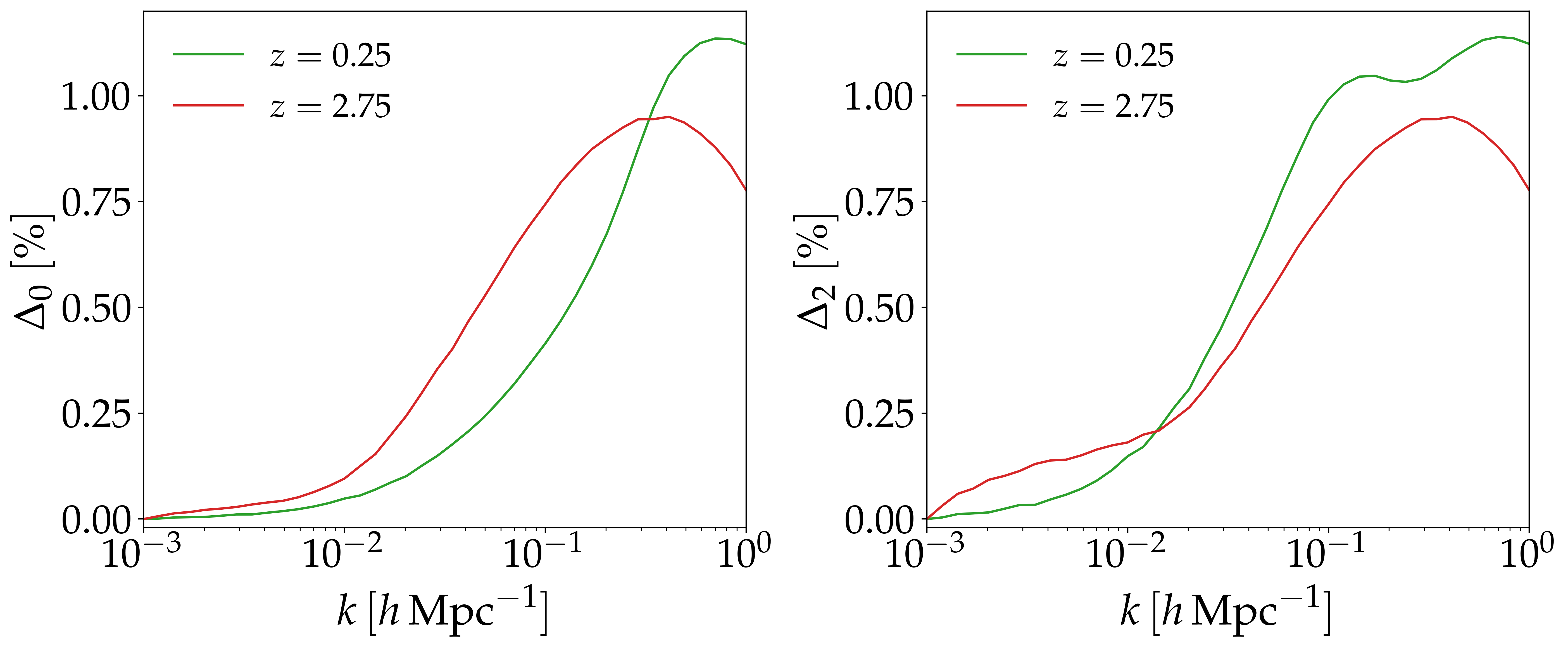}
    \caption{Percentage difference between multipoles of the 21cm IM power spectrum computed by considering $f(z)$ as a function of redshift only, with respect to the ones computed with the correct scale-dependent growth rate $f(z,k)$. We refer as $\Delta_0$ and $\Delta_2$ to the difference for the monopole and quadrupole, respectively. We show the cases of $z=0.25$ and $z=2.75$, which represent the limiting values in our analysis, for both the monopole (\textit{left panel}) and the quadrupole (\textit{right panel}).}
    \label{fig:P0_P2_diff}
\end{figure}
As discussed in \autoref{sec:modelling_21cm}, in cosmologies with massive neutrinos the growth rate $f$ depends on $k$ in addition to the redshift $z$. The growth rate is then defined as, 
\begin{equation}
    f(z, k) = -( 1 + z) \,\frac{d\log{D(z, k)}}{dz},
    \label{app:f_formula}
\end{equation}
where the growth factor $D(z,k)$ is obtained from the linear matter power spectrum as,
\begin{equation}
    D(z,k) = \biggl(\frac{P(z,k)}{P(z=0,k)}\biggr)^{\frac 1 2}.
    \label{app:growth_factor}
\end{equation}
In this work, whenever we use fixed nuisance parameters, we compute $f(z,k)$ directly from equation \eqref{app:f_formula}. However, when using nuisance parameters, as explained in \autoref{sec:likelihood}, $f(z,k)$ enters in one of the parameters combinations that is treated solely as a function of redshift in our analysis. Consequently, this approach neglects the scale dependence of $f(z,k)$. Including the scale dependence in the definition of the nuisance parameter is not feasible, as it would significantly increase the computational time required to run the MCMC analyses. \\
In \autoref{fig:f_cc}, we show the difference between the scale-invariant CAMB computation for $f(z)$, which can also be obtained via the fitting function $f(z) = \Omega_\mathrm{m}^{0.55}$, and the results from equation \eqref{app:growth_factor} for different values of $\Sigma m_{\mathrm{\nu}}$, at $z=0.25$ (left panel) and $z=2.75$ (right panel). \autoref{fig:P0_P2_diff} shows the corresponding error in the estimation of the 21cm IM power spectrum monopole and quadrupole when considering $f$ as a function of redshift only, for $z=0.25$ and $z=2.75$. In both cases, for the linear scales that are relevant to this work, the difference is $<1.0 \%$. {In this comparison we have fixed the matter density ($\Omega_{\rm m}$) to be the same across all the neutrino mass cases. }

We do not show here the signal-to-noise ratios for the 21cm IM power spectrum monopole and quadrupole, as we find very similar results to Figure 7 of \citep{Berti:2022ilk} and therefore we refer the reader to that reference. We find that the bin with the highest S/N is the lowest redshift bin, for which the S/N has a maximum value of $\sim 85$, implying that the errors on the multipoles observations are, in this redshift bin, always $\gtrsim 1.2 \%$. For the other redshift bins, the S/N is of the order of $\sim 50$ or lower, corresponding to a percentage error on the measurements of $\sim 2 \%$ or more. Overall, the uncertainties on the measurements are always larger but of comparable order to the error introduced by the neglecting the scale dependence of the growth rate. However, in a real survey, additional effects, such as residual foregrounds and systematic contamination, are expected to increase the measurement uncertainties, potentially making them dominant over the error introduced by neglecting the scale dependence of $f$. A more comprehensive investigation of these effects is beyond the scope of this work.

\section{Constraints on the full set of the analysis parameters}
\label{sec:full_constraints}
In this discussion, we present the constraints obtained with the synthetic data sets presented in the sections above on all cosmological parameters. We report results at $95\%$ confidence level for upper limits, while results with two-sided error bars refer to $68\%$ confidence intervals. In \autoref{tab:constraints_P21}, we present the constraints obtained with the 21cm IM data sets alone and in combination with CMB data, in \autoref{tab:constraints_cross_desi} and \autoref{tab:constraints_cross_euclid}, we present the constraints obtained with the SKAO$\times$DESI and SKAO$\times$Euclid cross-correlation data sets, respectively.

  \clearpage            
\begin{table}[p]     
\small   
\thispagestyle{empty}
\renewcommand{\arraystretch}{1.2}
\makebox[\textwidth][c]{%
  \begin{tabular}{p{2.5cm} C{1.3cm} C{2.3cm} C{2.3cm} C{2.7cm} C{2.7cm}}
    \toprule
    \multicolumn{6}{c}{21cm Intensity Mapping Constraints ($\Sigma m_\nu^{\rm fid} =0.06$ eV)} \\
    \toprule
    Parameters 
      & Fiducials  
      & $\hat{P}_0 + \hat{P}_2$
      & \shortstack{$\hat{P}_0 + \hat{P}_2$ \\ + nuis.} 
      & \shortstack{Planck \\ + $\hat{P}_0 + \hat{P}_2$}
      & \shortstack{Planck \\ +  $\hat{P}_0 + \hat{P}_2$\\  + nuis.} \\ 
    \toprule

    $\Omega_b h^2$ 
      & 0.02238 
      & $0.02234 \pm 0.00067$ 
      & $0.0229^{+0.0011}_{-0.00096}$ 
      & $0.02241 \pm 0.00010$
      & $0.02242 \pm 0.00011$ \\

    $\Omega_c h^2$ 
      & 0.11987 
      & $0.1198 \pm 0.0013$ 
      & $0.1211 \pm 0.0040$
      & $0.11972^{+0.00039}_{-0.00034}$ 
      & $0.11955^{+0.00054}_{-0.00044}$ \\

    $n_\mathrm{s}$
      & 0.96585 
      & $0.969 \pm 0.013$ 
      & $0.971 \pm 0.021$ 
      & $0.9665 \pm 0.0029$ 
      & $0.9665 \pm 0.0031$ \\

    $\ln (10^{10} A_s)$
      & 3.0444  
      &  $3.064^{+0.023}_{-0.041}$  
      &  $3.068 \pm 0.054$ 
      &  $3.0487^{+0.0042}_{-0.0080}$
      &  $3.046 \pm 0.013$\\

    $\tau$ 
      & 0.0543  
      & ---  
      & --- 
      &  $0.0565^{+0.0035}_{-0.0047}$
      &  $0.0554 \pm 0.0066$\\

    $100\theta_{\mathrm{MC}}$
      & 1.04091
      & $1.0425^{+0.0028}_{-0.0036}$
      & $1.0446^{+0.0035}_{-0.0043}$
      & $1.04096 \pm 0.00026$
      & $1.04094 \pm 0.00026$\\

    $\Sigma m_\nu$ (eV)
      & 0.06  
      & $<0.287$   
      & $<0.425$
      &  $<0.105$
      & $<0.126$\\

    $H_0$ (km/s/Mpc)
      & 67.41
      & $67.419 \pm 0.062$
      & $67.43 \pm 0.10$
      & $67.410 \pm 0.056$
      & $67.395 \pm 0.065$\\

    $\sigma_8$
      & 0.8109
      & $0.8082^{+0.0041}_{-0.0034}$
      & $0.796^{+0.024}_{-0.019}$
      & $0.8099 \pm 0.0018$
      & $0.8062^{+0.0071}_{-0.0060}$\\

    $\Omega_\mathrm{m}$
      & 0.3144
      & $0.3156 \pm 0.0033$
      & $0.321 \pm 0.010$
      & $0.3146 \pm 0.0010$
      & $0.3146 \pm 0.0012$\\
    \toprule
    
    \multicolumn{6}{c}{21cm Intensity Mapping Constraints ($\Sigma m_\nu^{\rm fid} =0.1$ eV)} \\
    \toprule
    Parameters 
      & Fiducials  
      & $\hat{P}_0 + \hat{P}_2$
      & \shortstack{$\hat{P}_0 + \hat{P}_2$ \\ + nuis.} 
      & \shortstack{Planck \\ + $\hat{P}_0 + \hat{P}_2$}
      & \shortstack{Planck \\ +  $\hat{P}_0 + \hat{P}_2$\\  + nuis.} \\ 
    \toprule
    $\Omega_b h^2$ 
      & 0.02235 
      & $0.02225 \pm 0.00069$ 
      & $0.0225 \pm 0.0011$ 
      & $0.02234 \pm 0.00011$ 
      & $0.02235 \pm 0.00011$\\

    $\Omega_c h^2$ 
      & 0.12032 
      & $0.1203 \pm 0.0014$ 
      & $0.1199 \pm 0.0047$
      & $0.12035 \pm 0.00043$ 
      & $0.12029^{+0.00065}_{-0.00052}$ \\

    $n_\mathrm{s}$
      & 0.96437 
      & $0.969 \pm 0.013$ 
      & $0.978 \pm 0.022$ 
      & $0.9642 \pm 0.0032$ 
      & $0.9644 \pm 0.0032$ \\

    $\ln (10^{10} A_s)$
      & 3.0426
      &  $3.068^{+0.022}_{-0.039}$  
      &  $3.067 \pm 0.058$ 
      &  $3.0424^{+0.0084}_{-0.012}$
      &  $3.044 \pm 0.014$\\

    $\tau$ 
      & 0.05326
      & ---  
      & $<0.512$ 
      &  $0.0530^{+0.0049}_{-0.0061}$
      &  $0.0539 \pm 0.0068$\\

    $100\theta_{\mathrm{MC}}$
      & 1.04082
      & $1.0428^{+0.0028}_{-0.0035}$
      & $1.0429 \pm 0.0051$
      & $1.04081 \pm 0.00027$
      & $1.04083 \pm 0.00027$\\

    $\Sigma m_\mathrm{\nu}$ (eV)
      & $0.1$
      & $<0.317$   
      & $<0.452$
      &  $<0.135$
      & $<0.151$\\

    $H_0$ (km/s/Mpc)
      & 66.91
      & $66.914 \pm 0.063$
      & $66.91 \pm 0.11$
      & $66.909 \pm 0.057$
      & $66.922 \pm 0.070$\\

    $\sigma_8$
      & 0.8042
      & $0.8017 \pm 0.0031$
      & $0.791^{+0.024}_{-0.019}$
      & $0.8044 \pm 0.0021$
      & $0.8047^{+0.0083}_{-0.0066}$\\

    $\Omega_\mathrm{m}$
      & 0.3210
      & $0.3224 \pm 0.0034$
      & $0.323 \pm 0.012$
      & $0.3211 \pm 0.0011$
      & $0.3209 \pm 0.0012$\\
    \toprule
    
    \multicolumn{6}{c}{21cm Intensity Mapping Constraints ($\Sigma m_\nu^{\rm fid} =0.4$ eV)} \\
    \toprule
    Parameters 
      & Fiducials  
      & $\hat{P}_0 + \hat{P}_2$
      & \shortstack{$\hat{P}_0 + \hat{P}_2$ \\ + nuis.} 
      & \shortstack{Planck \\ + $\hat{P}_0 + \hat{P}_2$}
      & \shortstack{Planck \\ +  $\hat{P}_0 + \hat{P}_2$\\  + nuis.} \\ 
    \toprule
    $\Omega_b h^2$ 
      & 0.02215
      & $0.02214 \pm 0.00077$        
      & $0.0218 \pm 0.0011$
      & $0.02216 \pm 0.00011$ 
      & $0.02217 \pm 0.00011$ \\

    $\Omega_c h^2$
      & 0.12209
      & $0.1225 \pm 0.0016$
      & $0.1214 \pm 0.0044$
      & $0.12215 \pm 0.00039$
      & $0.12206 \pm 0.00064$\\

    $n_\mathrm{s}$
      & 0.95916
      & $0.959^{+0.017}_{-0.022}$
      & $0.952 \pm 0.024$
      & $0.9590 \pm 0.0032$
      & $0.9588 \pm 0.0034$\\

    $\ln(10^{10} A_s)$
      & 3.0529
      & $3.055^{+0.047}_{-0.054}$
      & $3.048 \pm 0.064$
      & $3.052 \pm 0.010$
      & $3.055^{+0.014}_{-0.016}$\\

    $\tau$
      & 0.0561
      & $<0.502$
      & $<0.374$
      & $0.0557 \pm 0.0056$
      & $0.0572^{+0.0069}_{-0.0080}$\\

    $100\theta_{\mathrm{MC}}$
      & 1.04049
      & $1.0412^{+0.0040}_{-0.0045}$
      & $1.0389^{+0.0046}_{-0.0053}$
      & $1.04048 \pm 0.00027$
      & $1.04049 \pm 0.00028$\\

    $\Sigma m_\mathrm{\nu}$ (eV)
      & 0.4
      & $0.41^{+0.25}_{-0.24}$   
      & $0.34^{+0.32}_{-0.2}$   
      & $0.398 \pm 0.036$
      & $0.401^{+0.065}_{-0.066}$\\

    $H_0$ (km/s/Mpc)
      & 63.43
      & $63.432 \pm 0.068$
      & $63.41 \pm 0.11$
      & $63.430 \pm 0.062$
      & $63.443 \pm 0.083$\\

    $\sigma_8$
      & 0.74505
      & $0.7451^{+0.0021}_{-0.0026}$
      & $0.752 \pm 0.025$
      & $0.7453 \pm 0.0018$
      & $0.7454 \pm 0.0085$\\

    $\Omega_\mathrm{m}$
      & 0.3692
      & $0.3705 \pm 0.0044$
      & $0.365 \pm 0.013$
      & $0.3693 \pm 0.0013$
      & $0.3690 \pm 0.0015$\\

    \bottomrule
  \end{tabular}%
}
\caption{Marginalized constraints on cosmological parameters at 95\% confidence level for 21\,cm IM data (SKA) alone and in combination with CMB (Planck 2018), for three fiducial neutrino masses $\Sigma m_\nu = 0.06,\,0.1,\,0.4$\,eV.}
\label{tab:constraints_P21}
\end{table}

  \clearpage            
\begin{table}[p]     
\small   
\thispagestyle{empty}
\renewcommand{\arraystretch}{1.2}
\makebox[\textwidth][c]{%
  \begin{tabular}{p{2.5cm} C{1.3cm} C{2.3cm} C{2.3cm} C{2.7cm} C{2.7cm}}
    \toprule
    \multicolumn{6}{c}{SKA $\times$ DESI ($\Sigma m_\nu^{\rm fid} =0.06$) eV} \\
    \toprule
    Parameters 
      & Fiducials  
      & $\hat{P}_{21,g}^{\mathrm{DESI}}$
      & \shortstack{$\hat{P}_{21,g}^{\mathrm{DESI}}$ \\ + nuis.} 
      & \shortstack{Planck \\ + $\hat{P}_{21,g}^{\mathrm{DESI}}$}
      & \shortstack{Planck \\ +  $\hat{P}_{21,g}^{\mathrm{DESI}}$\\  + nuis.} \\ 
    \toprule

    $\Omega_b h^2$ 
      & 0.02238 
      & $0.0222^{+0.0016}_{-0.0018}$
      & $0.0261^{+0.0055}_{-0.0091}$ 
      & $0.02241 \pm 0.00011$ 
      & $0.02240 \pm 0.00012$\\

    $\Omega_c h^2$ 
      & 0.11987 
      & $0.1207^{+0.0047}_{-0.0053}$ 
      & $0.133^{+0.023}_{-0.031}$
      & $0.11973 \pm 0.00055$ 
      & $0.11971 \pm 0.00082$ \\

    $n_\mathrm{s}$
      & 0.96585 
      & $0.970 \pm 0.029$ 
      & $0.967^{+0.066}_{-0.084}$ 
      & $0.9660 \pm 0.0032$ 
      & $0.9659 \pm 0.0035$ \\

    $\ln (10^{10} A_s)$
      & 3.0444  
      & $3.131^{+0.060}_{-0.084}$  
      & $2.97^{+0.31}_{-0.23}$ 
      & $3.0518^{+0.0051}_{-0.010}$
      & $3.048 \pm 0.014$\\

    $\tau$ 
      & 0.0543  
      & ---  
      & --- 
      & $0.0580^{+0.0035}_{-0.0054}$
      & $0.0561 \pm 0.0071$\\

    $100\,\theta_{\mathrm{MC}}$
      & 1.04091
      & $1.0472 \pm 0.0060$
      & $1.047^{+0.014}_{-0.0067}$
      & $1.04092 \pm 0.00027$
      & $1.04091 \pm 0.00028$\\

    $\Sigma m_\nu$ (eV)
      & 0.06
      & ---
      & ---
      & $<0.116$
      & $<0.155$ \\
      
    $H_0$ (km/s/Mpc)
      & 67.41
      & $67.11 \pm 0.59$
      & $67.9^{+1.4}_{-1.8}$
      & $67.35 \pm 0.19$
      & $67.23 \pm 0.36$\\

    $\sigma_8$
      & 0.8109
      & $0.8038 \pm 0.0054$
      & $0.758 \pm 0.078$
      & $0.8102 \pm 0.0024$
      & $0.8053^{+0.0092}_{-0.0067}$\\

    $\Omega_\mathrm{m}$
      & 0.3144
      & $0.324^{+0.012}_{-0.014}$
      & $0.350^{+0.053}_{-0.066}$
      & $0.3152 \pm 0.0026$
      & $0.3167 \pm 0.0049$\\
    \toprule
    
    \multicolumn{6}{c}{SKA $\times$ DESI  ($\Sigma m_\nu^{\rm fid} =0.1$) eV} \\
    \toprule
    Parameters 
      & Fiducials  
      & $\hat{P}_{21,g}^{\mathrm{DESI}}$
      & \shortstack{$\hat{P}_{21,g}^{\mathrm{DESI}}$ \\ + nuis.} 
      & \shortstack{Planck \\ + $\hat{P}_{21,g}^{\mathrm{DESI}}$}
      & \shortstack{Planck \\ +  $\hat{P}_{21,g}^{\mathrm{DESI}}$\\  + nuis.} \\ 
    \toprule
 
    $\Omega_b h^2$ 
      & 0.02235 
      & $0.0222^{+0.0016}_{-0.0018}$ 
      & $0.0244^{+0.0057}_{-0.0082}$ 
      & $0.02234 \pm 0.00011$ 
      & $0.02236 \pm 0.00012$\\

    $\Omega_c h^2$ 
      & 0.12032 
      & $0.1211^{+0.0048}_{-0.0054}$ 
      & $0.128^{+0.024}_{-0.029}$
      & $0.12031 \pm 0.00059$ 
      & $0.12021 \pm 0.00083$ \\

    $n_\mathrm{s}$
      & 0.96437 
      & $0.970 \pm 0.029$
      & $0.979^{+0.065}_{-0.091}$
      & $0.9645 \pm 0.0035$
      & $0.9648 \pm 0.0035$ \\

    $\ln (10^{10} A_s)$
      & 3.0426
      & $3.123^{+0.063}_{-0.086}$  
      & $2.99^{+0.29}_{-0.23}$ 
      & $3.0434^{+0.0088}_{-0.016}$
      & $3.046 \pm 0.014$\\

    $\tau$ 
      & 0.05326
      & ---
      & ---
      & $0.0537^{+0.0048}_{-0.0080}$
      & $0.0548 \pm 0.0073$\\

    $100\,\theta_{\mathrm{MC}}$
      & 1.04082
      & $1.0464 \pm 0.0060$
      & $1.044^{+0.016}_{-0.0059}$
      & $1.04082 \pm 0.00028$
      & $1.04084 \pm 0.00028$\\

    $\Sigma m_\nu$ (eV)
      & 0.1
      & ---
      & ---
      & $<0.144$
      & $<0.177$ \\

    $H_0$ (km/s/Mpc)
      & 66.91
      & $66.64 \pm 0.59$
      & $67.1^{+1.4}_{-1.6}$
      & $66.91 \pm 0.19$
      & $66.89 \pm 0.37$\\
      
    $\sigma_8$
      & 0.8042
      & $0.7989 \pm 0.0054$
      & $0.757 \pm 0.076$
      & $0.8045 \pm 0.0025$
      & $0.803^{+0.011}_{-0.0074}$\\

    $\Omega_\mathrm{m}$
      & 0.3210
      & $0.330 \pm 0.014$
      & $0.343 \pm 0.059$
      & $0.3210 \pm 0.0027$
      & $0.3213 \pm 0.0050$\\
    \toprule
    
    \multicolumn{6}{c}{SKA $\times$ DESI ($\Sigma m_\nu^{\rm fid} =0.4$) eV} \\
    \toprule
    Parameters 
      & Fiducials  
      & $\hat{P}_{21,g}^{\mathrm{DESI}}$
      & \shortstack{$\hat{P}_{21,g}^{\mathrm{DESI}}$ \\ + nuis.} 
      & \shortstack{Planck \\ + $\hat{P}_{21,g}^{\mathrm{DESI}}$}
      & \shortstack{Planck \\ +  $\hat{P}_{21,g}^{\mathrm{DESI}}$\\  + nuis.} \\ 
    \toprule

    $\Omega_b h^2$ 
      & 0.02215
      & $0.0223 \pm 0.0018$
      & $0.0278^{+0.0075}_{-0.0094}$ 
      & $0.02216 \pm 0.00011$ 
      & $0.02214 \pm 0.00012$\\

    $\Omega_c h^2$
      & 0.12209
      & $0.1230 \pm 0.0056$
      & $0.142 \pm 0.028$
      & $0.12216 \pm 0.00057$
      & $0.12249 \pm 0.00093$\\

    $n_\mathrm{s}$
      & 0.95916
      & $0.954 \pm 0.030$
      & $0.928^{+0.054}_{-0.095}$
      & $0.9591 \pm 0.0035$
      & $0.9588 \pm 0.0036$\\

    $\ln(10^{10} A_s)$
      & 3.0529
      & $3.023 \pm 0.070$
      & $2.86^{+0.34}_{-0.25}$
      & $3.051^{+0.014}_{-0.015}$
      & $3.052 \pm 0.015$\\

    $\tau$
      & 0.0561
      & ---
      & $<0.472$
      & $0.0553 \pm 0.0075$
      & $0.0553^{+0.0069}_{-0.0078}$\\

    $100\,\theta_{\mathrm{MC}}$
      & 1.04049
      & $1.0395 \pm 0.0062$
      & $1.040^{+0.012}_{-0.0061}$
      & $1.04048 \pm 0.00029$
      & $1.04046 \pm 0.00028$\\

    $\Sigma m_\nu$ (eV)
      & 0.4
      & ---
      & ---
      & $0.396^{+0.052}_{-0.046}$
      & $0.35 \pm 0.12$\\
  
    $H_0$ (km/s/Mpc)
      & 63.43
      & $63.53 \pm 0.58$
      & $64.5^{+1.4}_{-1.7}$
      & $63.44 \pm 0.19$
      & $63.71 \pm 0.45$\\

    $\sigma_8$
      & 0.74505
      & $0.7471^{+0.0048}_{-0.0053}$
      & $0.719^{+0.094}_{-0.082}$
      & $0.7454 \pm 0.0022$
      & $0.757 \pm 0.014$\\

    $\Omega_\mathrm{m}$
      & 0.3692
      & $0.369 \pm 0.016$
      & $0.412 \pm 0.069$
      & $0.3691 \pm 0.0032$
      & $0.3656 \pm 0.0068$\\

    \bottomrule

  \end{tabular}%
}
\caption{Marginalized constraints on cosmological parameters at 95\% confidence level obtained by using the SKAO$\times$DESI synthetic data sets alone and in combination with CMB data, for three fiducial neutrino masses $\Sigma m_\nu = 0.06,\,0.1,\,0.4$\,eV.}
\label{tab:constraints_cross_desi}
\end{table}

\clearpage            
\begin{table}[p]     
\small   
\thispagestyle{empty}
\renewcommand{\arraystretch}{1.2}
\makebox[\textwidth][c]{%
  \begin{tabular}{p{2.5cm} C{1.3cm} C{2.3cm} C{2.3cm} C{2.7cm} C{2.7cm}}
    \toprule
    \multicolumn{6}{c}{SKA $\times$ Euclid ($\Sigma m_\nu^{\rm fid} =0.06$) eV} \\
    \toprule
    Parameters 
      & Fiducials  
      & $\hat{P}_{21,g}^{\mathrm{Euclid}}$
      & \shortstack{$\hat{P}_{21,g}^{\mathrm{Euclid}}$ \\ + nuis.} 
      & \shortstack{Planck \\ + $\hat{P}_{21,g}^{\mathrm{Euclid}}$}
      & \shortstack{Planck \\ +  $\hat{P}_{21,g}^{\mathrm{Euclid}}$\\  + nuis.} \\ 
    \toprule

    $\Omega_b h^2$
      & 0.02238
      & $0.0229^{+0.0018}_{-0.0021}$
      & $0.0298^{+0.0094}_{-0.012}$
      & $0.02240 \pm 0.00011$
      & $0.02240 \pm 0.00012$\\

    $\Omega_c h^2$
      & 0.11987
      & $0.1222 \pm 0.0064$
      & $0.143 \pm 0.032$
      & $0.11972 \pm 0.00054$
      & $0.11969 \pm 0.00081$\\

    $n_\mathrm{s}$
      & 0.96585
      & $0.973 \pm 0.028$
      & $0.957^{+0.066}_{-0.095}$
      & $0.9661 \pm 0.0033$
      & $0.9661 \pm 0.0035$\\

    $\ln(10^{10} A_s)$
      & 3.0444
      & $3.138^{+0.060}_{-0.083}$
      & $2.75^{+0.51}_{-0.39}$
      & $3.0515^{+0.0056}_{-0.011}$
      & $3.047 \pm 0.014$\\

    $\tau$
      & 0.0543
      & ---
      & $>0.413$
      & $0.0578^{+0.0038}_{-0.0058}$
      & $0.0560 \pm 0.0072$\\

    $100\,\theta_{\mathrm{MC}}$
      & 1.04091
      & $1.0474 \pm 0.0058$
      & $1.046^{+0.012}_{-0.0069}$
      & $1.04093 \pm 0.00027$
      & $1.04092 \pm 0.00028$\\

    $\Sigma m_\nu$ (eV)
      & 0.06
      & ---
      & ---
      & $<0.117$
      & $<0.156$\\

    $H_0$ (km/s/Mpc)
      & 67.41
      & $67.15 \pm 0.56$
      & $68.4 \pm 1.7$
      & $67.36 \pm 0.18$
      & $67.24 \pm 0.36$\\

    $\sigma_8$
      & 0.8109
      & $0.8063 \pm 0.0074$
      & $0.70 \pm 0.13$
      & $0.8100 \pm 0.0027$
      & $0.8051^{+0.0094}_{-0.0067}$\\

    $\Omega_\mathrm{m}$
      & 0.3144
      & $0.329^{+0.016}_{-0.018}$
      & $0.373 \pm 0.073$
      & $0.3151 \pm 0.0025$
      & $0.3165 \pm 0.0048$\\
    \toprule
    
    \multicolumn{6}{c}{SKA $\times$ Euclid  ($\Sigma m_\nu^{\rm fid} =0.1$) eV} \\
    \toprule
    Parameters 
      & Fiducials  
      & $\hat{P}_{21,g}^{\mathrm{Euclid}}$
      & \shortstack{$\hat{P}_{21,g}^{\mathrm{Euclid}}$ \\ + nuis.}  
      & \shortstack{Planck \\ + $\hat{P}_{21,g}^{\mathrm{Euclid}}$}
      & \shortstack{Planck \\ +  $\hat{P}_{21,g}^{\mathrm{Euclid}}$\\  + nuis.} \\ 
    \toprule

    $\Omega_b h^2$
      & 0.02235
      & $0.0226 \pm 0.0021$
      & $0.0260^{+0.0062}_{-0.0093}$
      & $0.02235 \pm 0.00011$
      & $0.02235 \pm 0.00012$\\

    $\Omega_c h^2$
      & 0.12032
      & $0.1221 \pm 0.0069$
      & $0.133^{+0.024}_{-0.031}$
      & $0.12033 \pm 0.00060$
      & $0.12022 \pm 0.00082$\\

    $n_\mathrm{s}$
      & 0.96437
      & $0.971 \pm 0.029$
      & $0.968^{+0.063}_{-0.085}$
      & $0.9643 \pm 0.0035$
      & $0.9648 \pm 0.0035$\\

    $\ln(10^{10} A_s)$
      & 3.0426
      & $3.122^{+0.061}_{-0.083}$
      & $2.85^{+0.48}_{-0.33}$
      & $3.0440^{+0.0093}_{-0.016}$
      & $3.046 \pm 0.014$\\

    $\tau$
      & 0.05326
      & ---
      & ---
      & $0.0538^{+0.0054}_{-0.0077}$
      & $0.0547 \pm 0.0072$\\

    $100\,\theta_{\mathrm{MC}}$
      & 1.04082
      & $1.0463 \pm 0.0059$
      & $1.045^{+0.013}_{-0.0058}$
      & $1.04083 \pm 0.00028$
      & $1.04083 \pm 0.00028$\\

    $\Sigma m_\nu$ (eV)
      & 0.1
      & ---
      & ---
      & $<0.145$
      & $<0.180$\\

    $H_0$ (km/s/Mpc)
      & 66.91
      & $66.67 \pm 0.55$
      & $67.2^{+1.2}_{-1.5}$
      & $66.90 \pm 0.18$
      & $66.87 \pm 0.38$\\

    $\sigma_8$
      & 0.8042
      & $0.8009 \pm 0.0078$
      & $0.72 \pm 0.13$
      & $0.8045 \pm 0.0030$
      & $0.803^{+0.011}_{-0.0076}$\\

    $\Omega_\mathrm{m}$
      & 0.3210
      & $0.333 \pm 0.019$
      & $0.357^{+0.061}_{-0.071}$
      & $0.3212 \pm 0.0026$
      & $0.3215 \pm 0.0051$\\
    \toprule
    
    \multicolumn{6}{c}{SKA $\times$ Euclid ($\Sigma m_\nu^{\rm fid} =0.4$) eV} \\
    \toprule
    Parameters 
      & Fiducials  
      & $\hat{P}_{21,g}^{\mathrm{Euclid}}$
      & \shortstack{$\hat{P}_{21,g}^{\mathrm{Euclid}}$ \\ + nuis.} 
      & \shortstack{Planck \\ + $\hat{P}_{21,g}^{\mathrm{Euclid}}$}
      & \shortstack{Planck \\ +  $\hat{P}_{21,g}^{\mathrm{Euclid}}$\\  + nuis.} \\ 
    \toprule

    $\Omega_b h^2$
      & 0.02215
      & $0.0226 \pm 0.0021$
      & $0.0272^{+0.0064}_{-0.011}$
      & $0.02216 \pm 0.00011$
      & $0.02215 \pm 0.00012$\\

    $\Omega_c h^2$
      & 0.12209
      & $0.1242^{+0.0069}_{-0.0078}$
      & $0.138^{+0.026}_{-0.035}$
      & $0.12216 \pm 0.00059$
      & $0.12247 \pm 0.00092$\\

    $n_\mathrm{s}$
      & 0.95916
      & $0.950 \pm 0.032$
      & $0.938^{+0.057}_{-0.095}$
      & $0.9592 \pm 0.0034$
      & $0.9588 \pm 0.0035$\\

    $\ln(10^{10} A_s)$
      & 3.0529
      & $3.023 \pm 0.070$
      & $2.76^{+0.53}_{-0.37}$
      & $3.052^{+0.013}_{-0.016}$
      & $3.052^{+0.014}_{-0.016}$\\

    $\tau$
      & 0.0561
      & ---
      & ---
      & $0.0557^{+0.0066}_{-0.0078}$
      & $0.0552^{+0.0070}_{-0.0079}$\\

    $100\,\theta_{\mathrm{MC}}$
      & 1.04049
      & $1.0398 \pm 0.0059$
      & $1.037^{+0.013}_{-0.0049}$
      & $1.04048 \pm 0.00027$
      & $1.04046^{+0.00029}_{-0.00029}$\\

    $\Sigma m_\nu$ (eV)
      & 0.4
      & ---
      & ---
      & $0.397^{+0.052}_{-0.049}$
      & $0.34 \pm 0.12$\\

    $H_0$ (km/s/Mpc)
      & 63.43
      & $63.55 \pm 0.57$
      & $64.4^{+1.2}_{-1.7}$
      & $63.43 \pm 0.20$
      & $63.78 \pm 0.48$\\

    $\sigma_8$
      & 0.74505
      & $0.7483 \pm 0.0081$
      & $0.69 \pm 0.13$
      & $0.7455 \pm 0.0027$
      & $0.758 \pm 0.014$\\

    $\Omega_\mathrm{m}$
      & 0.3692
      & $0.373^{+0.020}_{-0.024}$
      & $0.403^{+0.072}_{-0.086}$
      & $0.3693 \pm 0.0033$
      & $0.3647 \pm 0.0072$\\

    \bottomrule
  \end{tabular}%
}
\caption{Marginalized constraints on cosmological parameters at 95\% confidence level obtained by using the SKAO$\times$Euclid synthetic data sets alone and in combination with CMB data, for three fiducial neutrino masses $\Sigma m_\nu = 0.06,\,0.1,\,0.4$\,eV. }
\label{tab:constraints_cross_euclid}
\end{table}

\clearpage
\bibliographystyle{utcaps}
\bibliography{Bibliography}
\end{document}